\documentclass[12pt]{iopart}

\usepackage[lmargin=2.5cm,rmargin=2.5cm,tmargin=3cm,bmargin=2cm]{geometry}
\usepackage{amssymb}
\usepackage{graphicx}
\usepackage{graphics}
\usepackage{appendix}
\usepackage{multirow}
\usepackage{hhline}
\usepackage[normalem]{ulem}
\usepackage[usenames, dvipsnames]{color}
\usepackage{relsize}
\usepackage{color}
\usepackage{caption}
\usepackage{subcaption}
\usepackage{xspace}
\usepackage{cite}

\usepackage{hyperref}
\hypersetup{
    colorlinks=true,
    linkcolor=blue,
    filecolor=blue,      
    urlcolor=blue,
    citecolor=blue
}

\renewcommand{\sout}[1]{\unskip}

\newcommand{\red}[1]{#1}
\newcommand{\redd}[1]{#1}
\newcommand{\myWidth}{0.5\textwidth}
\newcommand{\halfWidth}{0.49\textwidth}
\newcommand{\pot}{90\deg}
\newcommand{\tB}{\theta_B}
\newcommand{\tp}{\theta_{\rm p}}
\newcommand{\tpol}{\theta_{\rm pol}}
\newcommand{\fpol}{\mathrm{f}_{\rm pol}}
\newcommand{\fcirc}{\mathrm{f}_{\rm circ}}
\newcommand{\tpmin}{\theta_{\rm p,min}}
\newcommand{\tpmax}{\theta_{\rm p,max}}
\newcommand{\Bt}{B_{\rm t}}
\newcommand{\Bp}{B_{\rm p}}
\newcommand{\vv}{\vec{v}}
\newcommand{\vB}{\vec{B}}
\newcommand{\vE}{\vec{E}}
\newcommand{\Rtan}{R_{\rm tan}}
\newcommand{\rtan}{r_{\rm tan}}

\renewcommand{\deg}{^\circ}
\newcommand{\Ghat}{\hat{G}}
\newcommand{\tpolhat}{\hat{\theta}_{\rm pol}}
\newcommand{\fpolhat}{\hat{\mathrm{f}}_{\rm pol}}
\newcommand{\tpc}{\theta_{\rm p,crit}}
\newcommand{\SOFT}{\textsc{Soft}\xspace}
\newcommand{\CODE}{\textsc{Code}\xspace}
\newcommand{\SOFTCODE}{\textsc{Soft+Code}\xspace}
\newcommand{\nRE}{n_{\rm RE}}
\newcommand{\Zeff}{Z_{\rm eff}}
\newcommand{\ehat}{\hat{e}}
\newcommand{\eqref}[1]{(\ref{#1})}
\newcommand{\dd}{\mathrm{d}}
\newcommand{\Psynch}{P_{\rm synch}}

\newcommand{\tBabs}{\vert\tB\vert}
\newcommand{\Zhat}{\hat{Z}}
\newcommand{\phihat}{\hat{\phi}}
\newcommand{\paren}[1]{\left(#1\right)}
\newcommand{\I}{\mathsf{I}}
\newcommand{\IMA}{I_{\rm MA}}

\newcommand{\LMSE}{L_{\rm MSE}}
\newcommand{\TkeV}{T_{\rm keV}}
\newcommand{\Ip}{I_{\rm p}}
\newcommand{\trad}{\tau_{\rm rad}}
\newcommand{\tcoll}{\tau_{\rm coll}}
\newcommand{\Bo}{B_0}
\newcommand{\Ec}{E_{\rm C}}

\begin{document}

\title[Polarized synchrotron emission from runaway electrons]{Experimental and synthetic measurements of polarized synchrotron emission from runaway electrons in Alcator~C-Mod}

\author{R.A.~Tinguely$^1$\footnote{Author to whom correspondence should be addressed: rating@mit.edu}, M.~Hoppe$^2$, R.S.~Granetz$^1$, R.T.~Mumgaard$^1$, and S.~Scott$^3$}

\address{$^1$ Plasma Science and Fusion Center, Massachusetts Institute of Technology, Cambridge, MA, USA \\
$^2$ Department of Physics, Chalmers University of Technology, G\"{o}teborg, Sweden \\
$^3$ Princeton Plasma Physics Laboratory, Princeton, NJ, USA}

\begin{abstract}
This paper presents the first experimental analysis of polarized synchrotron emission from relativistic runaway electrons (REs) in a tokamak plasma. Importantly, we show that the polarization information of synchrotron radiation can be used to diagnose spatially-localized RE pitch angle distributions. Synchrotron-producing REs were generated during low density, Ohmic, diverted plasma discharges in the Alcator C-Mod tokamak. The ten-channel Motional Stark Effect diagnostic was used to measure spatial profiles of the polarization angle $\tpol$ and the fraction $\fpol$ of detected light that was linearly-polarized. Spatial transitions in $\tpol$ of $\pot$---from horizontal to vertical polarization and vice versa---are observed in experimental data and are well-explained by the gyro-motion of REs and high directionality of synchrotron radiation. Polarized synchrotron emission is modeled with the synthetic diagnostic \SOFT; its output Green's (or detector response) functions reveal a critical RE pitch angle at which $\tpol$ flips by 90$\deg$ and $\fpol$ is minimal. Using \SOFT, we determine the dominant RE pitch angle which reproduces measured $\tpol$ and $\fpol$ values. The spatiotemporal evolutions of $\tpol$ and $\fpol$ are explored in detail for one C-Mod discharge. For channels viewing REs near the magnetic axis and flux surfaces $q$~=~1 and 4/3, disagreements between synthetic and experimental signals suggest that the sawtooth instability may be influencing RE dynamics. Furthermore, other sources of pitch angle scattering, not considered in this analysis, could help explain discrepancies between simulation and experiment.


\end{abstract}




\section{Introduction}\label{sec:introduction}

The gyration of a highly energetic charged particle in a background magnetic field $\vB$ produces synchrotron radiation~\cite{schwinger1949}, the relativistic extension of cyclotron radiation. Synchrotron light is directed primarily along the particle's velocity $\vv$, resulting in a forward ``beam'' of emission with angular width $\sim$1/$\gamma$, where $\gamma = 1/\sqrt{1-v^2/c^2}$ is the relativistic factor. It is well known that the classical polarization of electromagnetic radiation is mostly parallel to the charged particle's acceleration, or instantaneous radius of curvature in the case of gyro-motion. The first calculation of the polarization of synchrotron radiation was performed by Westfold in 1959~\cite{westfold1959}; this was motivated by experimental observations of highly polarized light from the Crab Nebula~\cite{vashakidze1954,dombrovsky1954} and the hypothesis that ultra-relativistic electrons moving in a magnetic field were the source~\cite{shklovskii1953}. 

In tokamak plasmas, synchrotron radiation is observed from relativistic ``runaway'' electrons (REs), which can be generated by sufficiently strong electric fields during plasma current ramp-up, low density discharges, or major disruptions. Many past works have studied the spectra \cite{yu2013,zhou2014,popovic2016,esposito2017,tinguely2018nf} and images\footnote{See table~1 in \cite{tinguely2018ppcf} for a more extensive list of references.} \cite{yu2013,zhou2014,paz-soldan2014,tinguely2018ppcf,hoppe2018d3d} of synchrotron emission, exploring RE energy evolution and spatiotemporal dynamics. This paper presents the first analysis of experimentally-measured \emph{polarized} synchrotron radiation from REs in tokamak plasmas. The first theoretical analysis of polarization properties and synthetic measurements of RE synchrotron radiation in a tokamak was performed by Sobolev in~\cite{sobolev2013}, which suggested that polarization information could be used to diagnose RE beams. As will be discussed, polarization measurements of synchrotron emission can provide insight into the distribution of RE pitch angles $\tp$. Thus, synchrotron polarization can help better constrain kinetic models of RE evolution and investigate mechanisms for increased pitch angle scattering, such as wave-particle instabilities \cite{liu2018} and high-Z material injection \cite{hesslow2017,hesslow2018}. Ultimately, understanding these RE dynamics and power loss mechanisms---e.g. radiated synchrotron power which increases with pitch angle, $\Psynch \propto \sin^2\tp$---can inform RE avoidance and mitigation strategies for future devices such as ITER \cite{lehnen2009} and SPARC \cite{greenwald2018}.

The outline of the rest of the paper is as follows: Section~\ref{sec:experiment} discusses the experimental setup and measurements of polarized synchrotron emission. Section~\ref{sec:modeling} describes the modeling of polarized synchrotron light and calculations of synthetic signals. In section~\ref{sec:comparison}, polarization data from one Alcator C-Mod discharge is explored in detail and compared to synthetic models of measurements. Finally, a discussion and summary is presented in section~\ref{sec:summary}.


\section{Experimental setup}\label{sec:experiment}

These RE studies were performed on the Alcator C-Mod tokamak \cite{marmar2009}, a compact device ($R_0$~=~68~cm, $a$~=~22~cm) with a toroidal magnetic field up to 8~T, including operation at $B_0$~=~5.3~T, as planned for ITER. Due to the high field, a significant portion of the synchrotron emission spectrum---that is produced by REs with energies of a few tens of MeV---falls in the visible wavelength range. In C-Mod, these synchrotron-producing REs are not observed after plasma disruptions, likely due to the fast break-up of the magnetic topology~\cite{marmar2009,izzo2011}; therefore, to study these effects, REs are purposefully generated during the flattop plasma current of low-density, Ohmic discharges. The plasmas explored in this study are also elongated and diverted. Time traces of one sample discharge, explored in detail in section~\ref{sec:comparison}, are shown in figure~\ref{fig:params}.

\begin{figure}[h!]
    \centering
    \includegraphics[width=\myWidth]{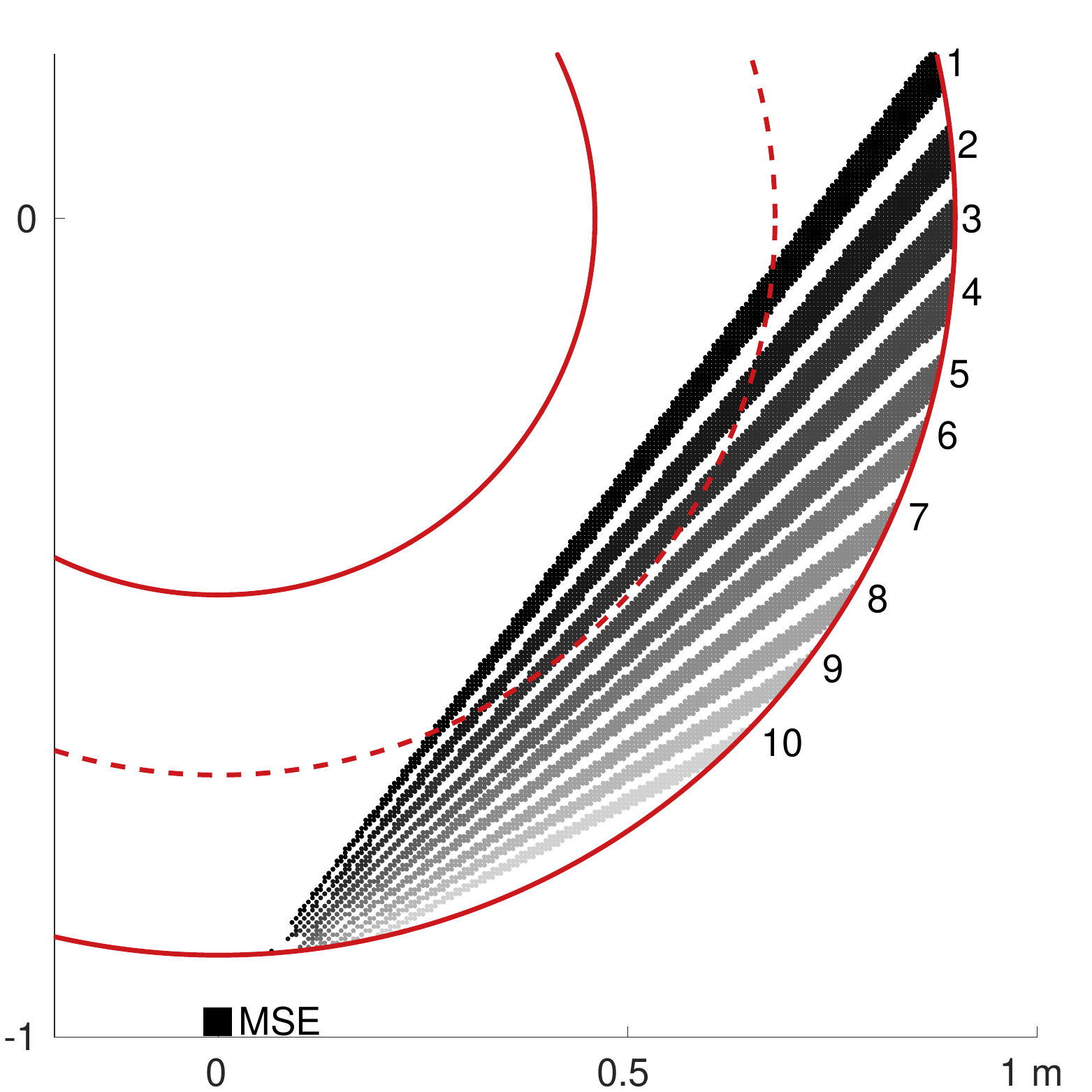}
    \caption{A top-down schematic of the ten-channel MSE diagnostic (black box) and its fields-of-view. The plasma boundary (solid) and magnetic axis (dotted) are overlaid.}
    \label{fig:mse}
\end{figure}

In C-Mod, polarization information of synchrotron radiation is gathered by the ten-channel Motional Stark Effect (MSE) diagnostic~\cite{mumgaard2016} which accepts light within a narrow wavelength band, $\Delta \lambda \approx$~0.8~nm, centered at $\lambda \approx$~660~nm in the visible range. During routine tokamak operation, the MSE system serves a completely separate purpose by measuring a different source of polarized light: line radiation resulting from electron transitions between Stark-shifted atomic (hydrogen) energy levels. This measurement requires neutral atoms within an injected diagnostic neutral hydrogen beam (DNB). The field-of-view (FOV) of each MSE channel is small (see figure~\ref{fig:mse}), with a total opening angle of $2\alpha \approx 1.7\deg$, meaning that each channel's measurement of emission from the DNB is radially-localized. The polarization angle $\tpol$ of the detected isotropic emission indicates the pitch of the local magnetic field $\tan \tB = \Bp/\Bt$, where $\Bp$ and $\Bt$ are the local poloidal and toroidal fields, respectively. The DNB was not in use during any of the RE experiments reported in this paper. \red{Furthermore, the contribution of polarized light from the background plasma was found to be insignificant ($<5\%$) based on MSE measurements when REs were not present.} Therefore, \red{we are confident that} the detected signal is \red{in fact} dominated by synchrotron emission\red{, with signal-to-noise ratios ranging from $\sim$20-100}. In addition, the only times considered in this study are those when lower hybrid current drive was disengaged.

\begin{figure}[h!]
    \centering
    \begin{subfigure}{\halfWidth}
        \centering
        \includegraphics[width=\textwidth]{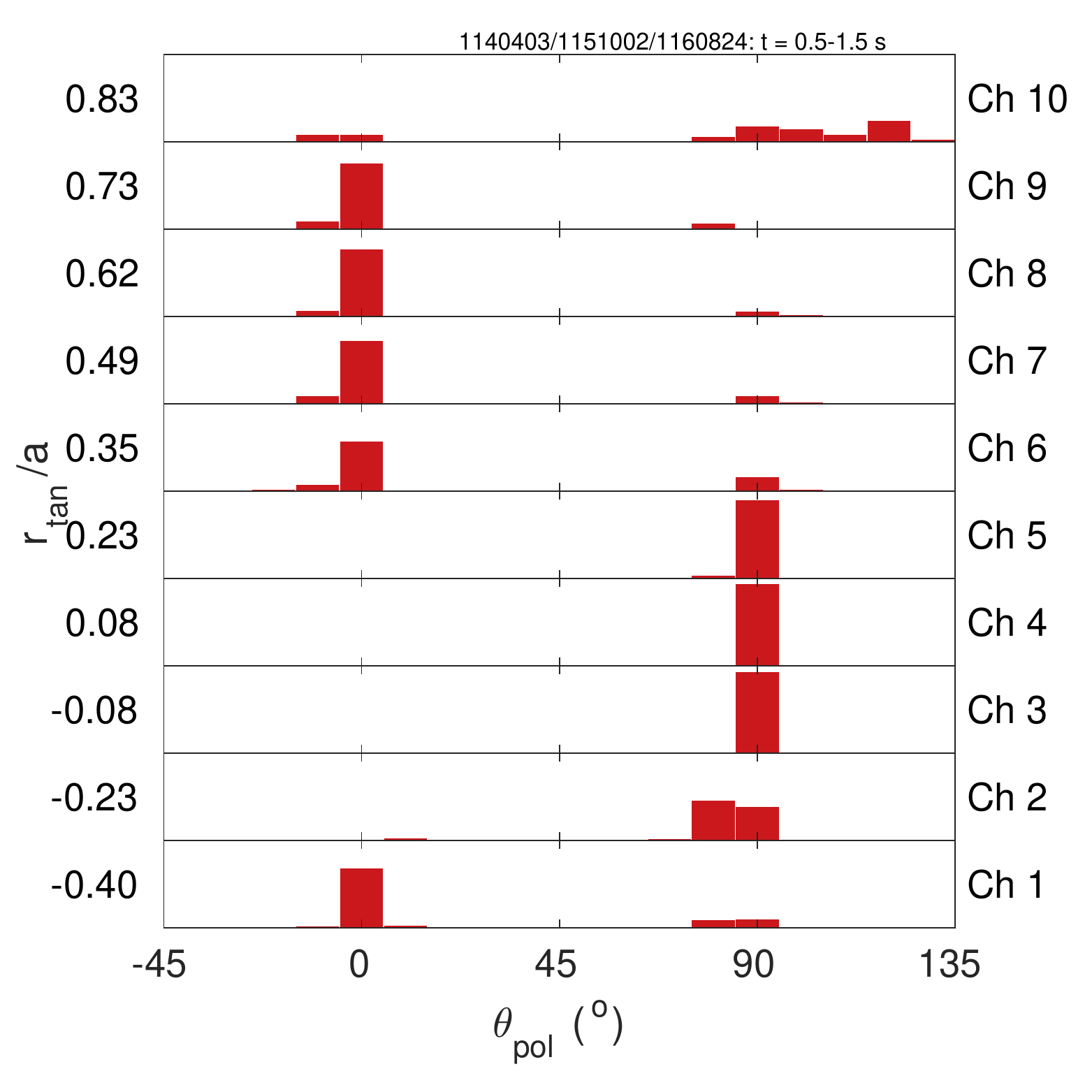}
        \caption{}
        \label{fig:hist_pa}
    \end{subfigure}
    \begin{subfigure}{\halfWidth}
        \centering
        \includegraphics[width=\textwidth]{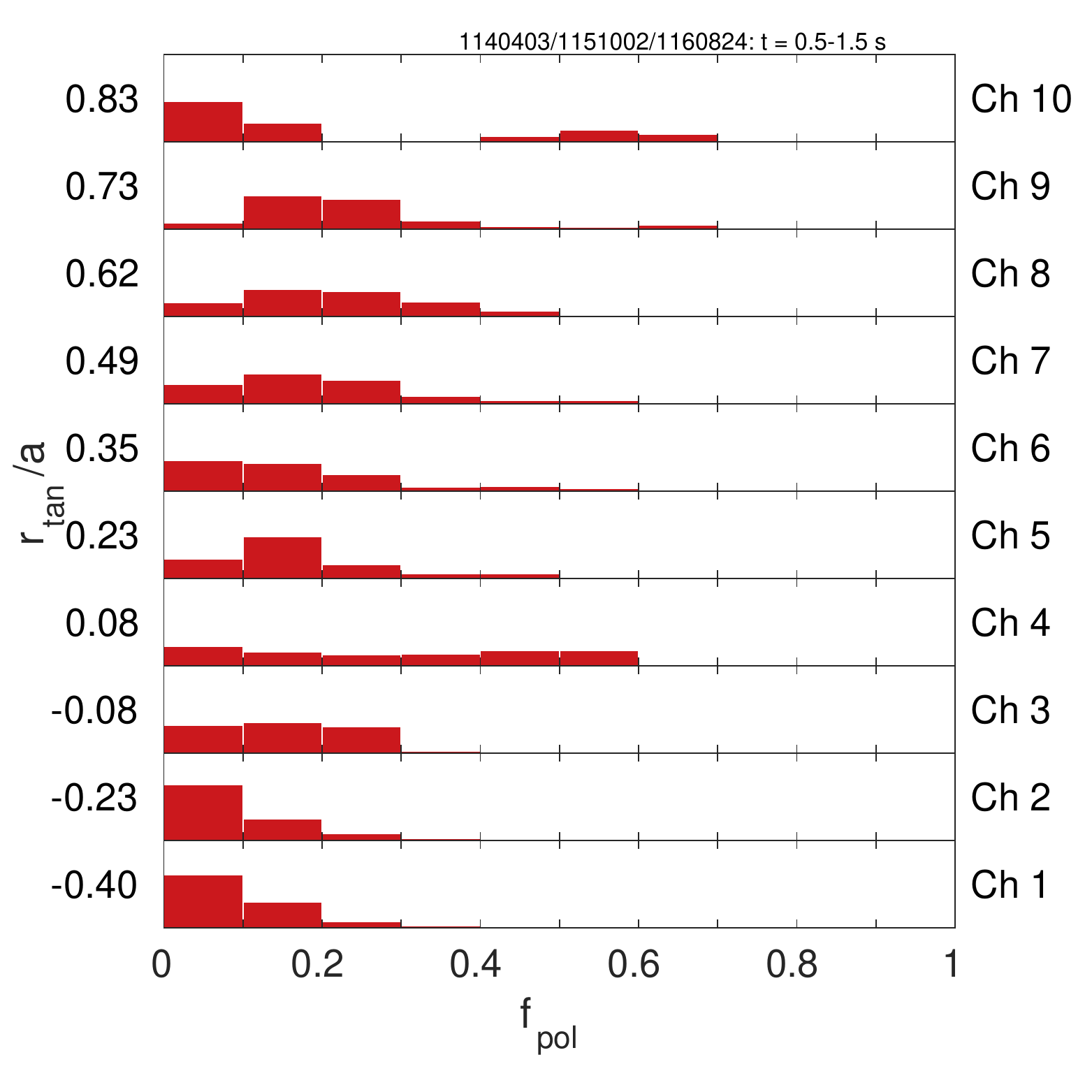}
        \caption{}
        \label{fig:hist_pf}
    \end{subfigure}
    \caption{Histograms of polarization (a) angle $\tpol$ and (b) fraction $\fpol$ for each channel, denoted by the normalized tangency radius $\rtan/a$. Bin widths are $\Delta \tpol$~=~$10^\circ$ and $\Delta\fpol$~=~0.1. For each channel, the vertical axis spans 0 to 1, and all bar heights (probabilities) sum to 1. Data are from 28 discharges and over 1000 time-slices during the plasma current flattop ($t$~=~0.5-1.5~s).}
\end{figure}

A top-down schematic of the MSE diagnostic and its ten FOV is depicted in figure~\ref{fig:mse}. A list of detector specifications are given in table~\ref{tab:MSE}. Although each channel makes a volume-integrated measurement within its FOV, the tangency major radius, or impact parameter $\Rtan$, is used to identify the major radius at which each line-of-sight (LOS) is orthogonal to the local major radial vector $\hat{R}$. The tangency minor radius is defined as $\rtan = \Rtan - R_0$ and falls in the range $\rtan \in [-a,a]$. See figure~\ref{fig:hist_pa} for the values of the \emph{normalized} tangency radius $\rtan/a$ of each channel and figure~\ref{fig:topdown} for an illustration. When RE pitch angles are smaller than the local magnetic pitch angle (discussed further in section~\ref{sec:heuristic}), it is expected that most detected synchrotron emission will come from REs located near $\Rtan$. Note that the MSE diagnostic is situated slightly above the midplane; therefore, the channels have a small downward-viewing orientation---i.e. an inclination $\delta < 0$---to intersect the midplane trajectory of the DNB. 

\begin{table}[h!]
	\centering
	\caption{Specifications of the MSE diagnostic in Alcator C-Mod. See figure~\ref{fig:hist_pa} for the normalized tangency radius $\rtan/a$ of each channel.}
	\label{tab:MSE}
	\begin{tabular}{l c}
		\hline
		Specification & Value \\
		\hline
		Major radial position & $R\approx$~98.1~cm \\
		Vertical position & $Z\approx$~2.9~cm \\
		Wavelength range & $\lambda \in 660\pm0.4$~nm \\
		Time resolution & $\Delta t \approx$~1~ms \\
		Inclination & $\delta \approx -3\deg$ \\
		FOV opening angle & $2 \alpha \approx$~1.7$^\circ$ \\
		\hline
	\end{tabular}
\end{table}

The MSE system measures a spatial profile of synchrotron polarization information. Each channel records the total intensity of detected light $I$, intensity of linearly-polarized light $L$, and linear polarization angle $\tpol$. Because the absolute value of sensitivity varies \sout{between} \red{among the} channels, it is most useful to consider the \emph{degree} of linear polarization, i.e. the fraction of detected light that is linearly-polarized: $\fpol = L/I \in [0,1]$. These polarization fraction measurements are calibrated, with estimated uncertainties of $\sim$10\%. The polarization angle measurement has lower errors of less than a degree. Experimental measurements of $\tpol$ and $\fpol$, and their comparisons with synthetic models, will be the focus of this paper, although qualitative trends of $L$ will be explored for one discharge in section~\ref{sec:comparison}.

In total, polarization data of RE synchrotron emission were collected from 28 plasma discharges\footnote{\red{The discharge numbers are 2, 3, 6, 7, 14, 16, 17, 19, 20, 23-26, 30, and 34 for experiment~\#1140403; 15 and 18-22 for experiment~\#1151002; and 20, 21, and 23-27 for experiment~\#1160824.}} at over one-thousand time points. Plasma parameters for these discharges span line-averaged electron densities $\bar{n}_e \approx$~0.2-1.0~$\times$~10$^{20}$~m$^{-3}$, central electron temperatures $T_{e0} \approx$~1-5~keV, plasma currents $\Ip \approx$~0.4-1.0~kA, and on-axis magnetic fields $\Bo \approx$~2-6~T. Ranges of RE-relevant parameters include the ratios of the electric to critical field \cite{connor1975} $E/\Ec \approx$~3-15 and collisional to synchrotron radiation timescales $\tcoll/\trad >$~0.1; these indicate the relative strengths of the electric force and radiation damping to collisional friction, respectively. Histograms of measured $\tpol$ and $\fpol$ are shown in figures~\ref{fig:hist_pa} and \ref{fig:hist_pf}, respectively. The distributions of $\tpol$ are well-localized for most channels, and an interesting spatial trend is observed: At the innermost radius (channel~1, $\rtan/a = -0.4$), $\tpol \approx 0\deg$ indicates \emph{horizontally}-polarized synchrotron light.\footnote{Note that this convention of measuring $\tpol$ upward from the midplane is \emph{opposite} that normally used for MSE measurements on C-Mod, where $\tpol$ is instead measured downward from the vertical axis.} For $\vert\rtan/a\vert \leq 0.23$, i.e. near the plasma center, measurements of $\tpol \approx 90\deg$ indicate \emph{vertical} polarization of the synchrotron radiation. Beyond $\rtan/a \geq 0.35$, the polarization angle flips back to $\tpol \approx 0\deg$, although the outermost radius, $\rtan/a = 0.83$, shows a more evenly-distributed range of measured $\tpol$, possibly due to a lack of (synchrotron-emitting) REs at the plasma edge. This $\pot$ transition in space was actually predicted by Sobolev~\cite{sobolev2013} and is explained more intuitively in section~\ref{sec:heuristic}. Also, a range of $\tpol \in [-45\deg,135\deg)$ is considered because there is a $180\deg$ degeneracy in $\tpol$. 

Compared to $\tpol$, experimental measurements of $\fpol$ do not exhibit as clear a spatial trend. Still, there are some features to note in figure~\ref{fig:hist_pf}: A peak in $\fpol$ ($\sim$0.6) is often observed on channel~4 which has a FOV near the magnetic axis ($\rtan/a = 0.08$). There are also instances of $\fpol \approx$~0.6-0.7 near the plasma edge (channels~9-10), but these may not always be dominated by synchrotron light. Values of $\fpol \geq 0.7$ are not recorded\red{; the reason for this is not clear since unpolarized background plasma light is expected to contribute less than 5\% to the total signal. One possible explanation is that reflections of synchrotron light from C-Mod's metal wall contribute to the fraction of emission which \emph{appears} to be not-linearly-polarized.} \sout{one plausible explanation for this is that light from the plasma and/or reflections from C-Mod's metal wall contributed to the fraction of emission which is not linearly polarized. While this could mean that up to 30\% of the detected light is not synchrotron emission, the good agreement seen in the comparisons of synthetic and experimental $\tpol$ signals---discussed in section~\ref{sec:comparison}---indicates that any extra light is likely mostly unpolarized. If the background light is instead negligible compared to synchrotron radiation, this would imply that} \red{Even so, it could be that} the polarization fraction of the observed synchrotron emission is truly $\fpol \leq 0.7$, which is achievable according to the analysis in section~\ref{sec:comparison}. \red{Finally,} \sout{Furthermore,} note in figure~\ref{fig:hist_pf} that the \emph{mean} $\fpol$ values of channels~7-9 ($\rtan/a \in [0.49, 0.73]$) are higher than the those for channels~6 and 10 ($\rtan/a$~=~0.35 and 0.83). This spatial ``bump'' of mean $\fpol$ values for $\rtan/a \in [0.23, 0.83]$ (channels~5-10) \sout{means} \red{implies that} there is usually a local minimum in $\fpol$ around $\rtan/a \approx$~0.23-0.35. This is a slightly non-intuitive result since we might expect the intensity of (linearly-polarized) light to increase from the plasma edge to the center.

\section{Modeling polarization measurements}\label{sec:modeling}

This section describes the modeling of polarized synchrotron light and accompanying synthetic diagnostic simulations. This is needed not only to understand the trends seen in experimental data, but also to validate theoretical models of RE phase space evolution, which should reproduce experimental results. A brief introduction of the polarization of synchrotron emission is provided in section~\ref{sec:synchrotron}. Section~\ref{sec:heuristic} gives a heuristic picture of polarized synchrotron emission and a more physically-intuitive explanation for the $\pot$ transition observed in the polarization angle $\tpol$. Section~\ref{sec:SOFT} describes the calculation and implementation of polarization in the versatile synthetic diagnostic \emph{Synchrotron-detecting Orbit Following Toolkit} (\SOFT)~\cite{hoppe2018}. Then, in section~\ref{sec:response}, the detector response functions for synthetic measurements of $\tpol$ and $\fpol$ are shown to be powerful tools for interrogating the RE pitch angle distribution.

\subsection{Polarization of synchrotron radiation}\label{sec:synchrotron}



The synchrotron radiation electric field vector can be decomposed into two parts~\cite{jackson1999}
\begin{equation}\label{eq:Edecomposed}
    \vec{E} = \ehat_{\perp} E_\perp + i\ehat_\parallel E_\parallel,
\end{equation}
\redd{where $\ehat_\parallel$ is a unit vector in the direction of acceleration, and $\ehat_\perp\propto\vec{n}\times\ehat_\parallel$, with $\vec{n}$ a unit vector directed from the electron toward the observer.} For synchrotron radiation, which stems from the rapid gyro-motion around magnetic field lines, the acceleration vector is determined by the Lorentz force, meaning that $\ehat_\parallel\propto\vec{v}\times\vec{B}$, with $\vec{v}$ denoting the electron velocity and $\vec{B}$ the local magnetic field vector. It can be shown that for highly relativistic electrons, $E_\parallel\gg E_\perp$, so that the radiation is mainly linearly polarized in the $\ehat_\parallel$ direction~\cite{ginzburg1969}.


One convenient and complete way of expressing the polarization of electromagnetic radiation is using the four Stokes parameters~\cite{stokes1851}
\begin{eqnarray*}
    I &= \left| E_{\perp} \right|^2 + \left| E_{\parallel} \right|^2,\\
    Q &= \left| E_{\perp} \right|^2 - \left| E_{\parallel} \right|^2,\\
    U &= 2\mathrm{Re}\left( E_{\perp} E_{\parallel}^* \right),\\
    V &= -2\mathrm{Im}\left( E_{\perp} E_{\parallel}^* \right),
\end{eqnarray*}
where $E_\perp^*$ and $E_\parallel^*$ are the complex-conjugates of $E_\perp$ and $E_\parallel$. The Stokes parameters from an ensemble of particles, with different radiation fields $\vec{E}$, can be determined simply by a linear combination of individual Stokes \emph{vectors} $[I,Q,U,V]$. Calculating our quantities of interest from the Stokes parameters is straightforward: $I$ is just the total intensity. The intensity of \emph{linearly}-polarized light is $L = \sqrt{Q^2+U^2}$. Then the degree of linear polarization is $\fpol = L/I$, as stated previously. The degree of \emph{circular} polarization is given by $\fcirc = V/I$, although synchrotron emission from highly relativistic electrons is \emph{not} expected to have significant circular polarization~\cite{korchakov1962}. Finally, the polarization angle is $2\tpol = \arctan(U/Q)$. 


\subsection{A heuristic picture}\label{sec:heuristic}

\begin{figure}[h!]
    \centering
    \begin{subfigure}{0.37\textwidth}
        \centering
        \includegraphics[width=\textwidth]{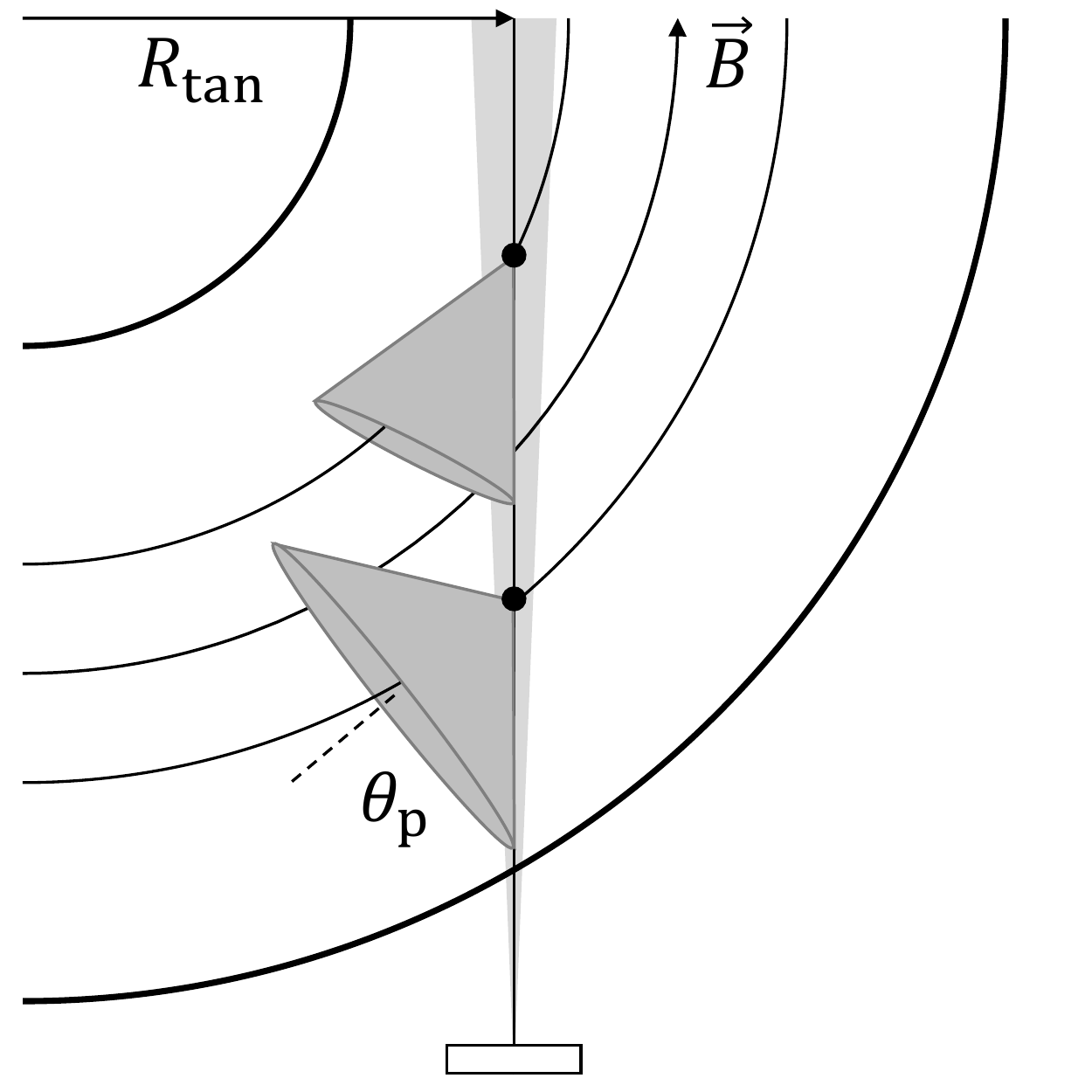}
        \caption{}
        \label{fig:topdown}
    \end{subfigure}
    \begin{subfigure}{0.62\textwidth}
        \centering
        \includegraphics[width=\textwidth]{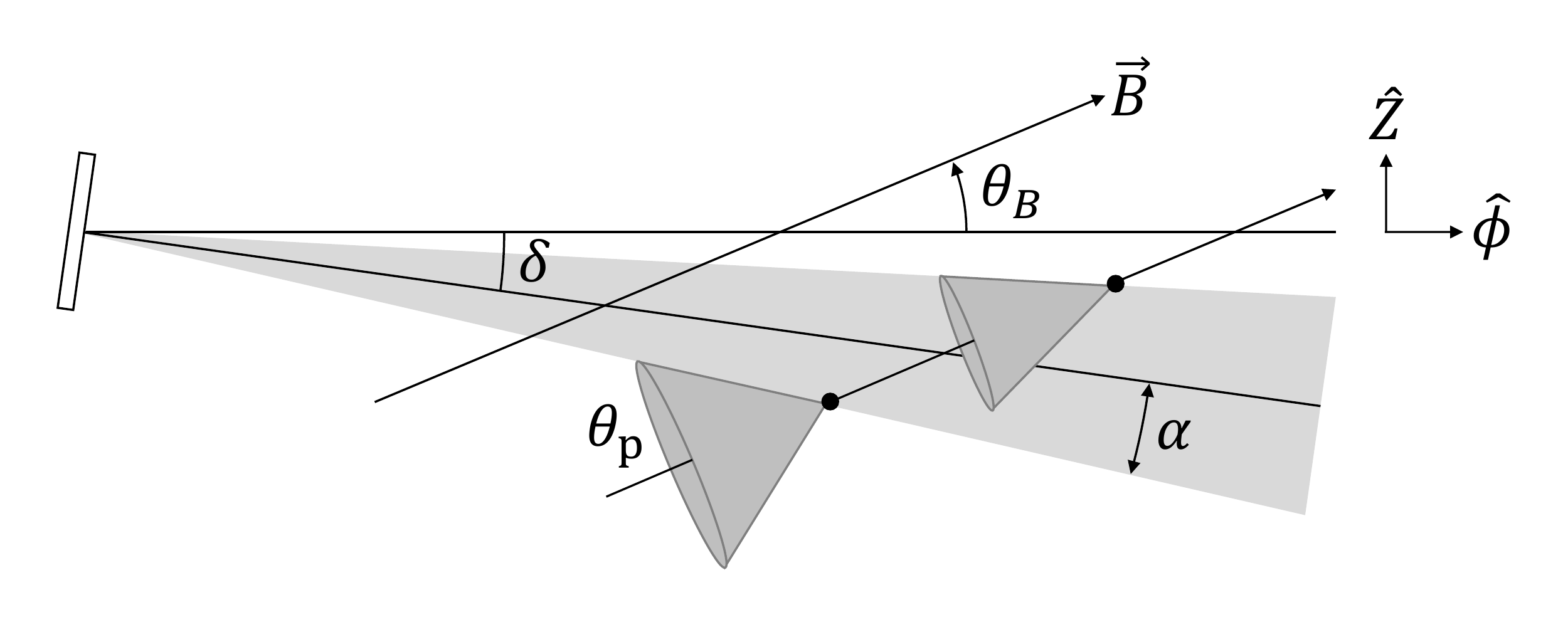}
        \caption{}
        \label{fig:toymodel}
    \end{subfigure}
    \caption{(a) A horizontal cross-section: A detector (bottom), with line-of-sight tangency radius $\Rtan$, views two REs (black dots) moving in approximately circular orbits with different pitch angles $\tp$ at two radii. (b) A vertical cross-section: A detector (left), with inclination $\delta < 0$ and opening half-angle $\alpha > 0$, views REs with pitch angles $\tpmin = \vert\tB - \delta\vert - \alpha$ (upper) and $\tpmax = \vert\tB - \delta\vert + \alpha$ (lower) in a magnetic field $\vB$ with local pitch $\tan\tB > 0$. The vertical axis is $\Zhat$, and \emph{local} toroidal vector is $\phihat$. (Not to-scale.)}
\end{figure}

In the guiding-center picture (schematically represented in figures~\ref{fig:topdown} and \ref{fig:toymodel}), a RE travels \emph{anti-parallel}\footnote{In Alcator C-Mod, the plasma current and toroidal magnetic field are usually \emph{parallel}, so that a strong electric field drives REs in the direction \emph{anti-parallel} to $\vB$.} to $\vB$, emitting a ``cone'' of synchrotron radiation in its forward direction with an opening half-angle equal to the RE pitch angle $\tp$. Here, the pitch angle is defined by $\tan \tp = v_\perp/v_\parallel$, where $v_\parallel$ and $v_\perp$ are the components of the RE velocity $\vv$ parallel and perpendicular to $-\vB$, respectively. Of course, this synchrotron radiation is only \emph{observed} if the cone (or, equivalently, $\vv$) is directed within a detector's FOV toward the detector's aperture.

On the plasma midplane, the pitch of $\vB$ is simply given by $\tan\tB = \Bp/\Bt$. In tokamaks, $\Bt$ is sufficiently larger than $\Bp$ (typically $\Bt/\Bp \approx 10$), so that $\vB$ is approximately horizontal. In fact, for realistic safety factor profiles---e.g. $q(0) \approx 1$ and $q_{95} > 3$---the magnetic field pitch is $\tBabs \leq$~0.2~rad. Therefore, in order for a detector to measure \emph{horizontally}-polarized synchrotron emission ($\tpol \approx 0\deg$), it must view the ``top'' and ``bottom'' of the emission cone. That is, $\vv$ should lie in the \emph{vertical} plane so that $\vE$ is horizontal since, as noted in section~\ref{sec:synchrotron}, $\vE$ is approximately proportional to $\vv \times \vB$. Conversely, to observe \emph{vertically}-polarized light ($\tpol \approx 90\deg$), a detector should view the ``sides'' of the synchrotron cone, when $\vv$ lies in the \emph{horizontal} plane.\footnote{\redd{Interestingly, synchrotron light directed along the side(s) of the emission cone is actually radiated by a RE located at the top/bottom of its gyro-orbit, and vice versa.}}

The clear spatial trends in the data of figures~\ref{fig:hist_pa} and \ref{fig:hist_pf}---$\tpol$, in particular---indicate a strong dependence of the experimental measurements on geometry. To give the reader a better intuition for \emph{why} a $\pot$ transition is observed across the plasma, consider a detector situated on the midplane with a tangency radius $\Rtan$, inclination $\delta$, and opening half-angle $\alpha$. Imagine a midplane cross-section of the plasma and detector view, as shown in figure~\ref{fig:topdown}. In this simplistic model, the detector will receive synchrotron light from REs at all $R \gtrsim \Rtan$, with the observable $\tp$ increasing as $R$ increases. For small $\alpha$ and $\delta$, radiation from REs with $\tp > \tB$ (corresponding to REs at $R > \Rtan$) is primarily \emph{vertically}-polarized. That is, the velocity $\vv$ of these REs must lie approximately in the midplane for light to reach the detector, and only the ``sides'' of the cone are observed. 

However, for REs located close to $\Rtan$ with $\tp \approx \tB$, the local magnetic field pitch must be taken into account. A vertical cross-section near $\Rtan$ is shown in figure~\ref{fig:toymodel}. The range of RE pitch angles for which the detector will measure synchrotron light is bounded by $\tpmin \leq \tp \leq \tpmax$, where
    \begin{equation}
        \theta_{\rm p,min}^{\rm p,max} = \max\paren{\vert\tB - \delta\vert \pm \alpha, \, 0}.
        \label{eq:heuristicCalc}
    \end{equation}
Here, the superscript and subscript correspond to the $\pm$ sign. Both $\tB$ and $\delta$ are measured positively counter-clockwise from $\phihat$ toward $\Zhat$ (see figure~\ref{fig:toymodel}), and both $\alpha > 0$ and $\tp > 0$ are assumed. Since $\alpha$ and $\delta$ are small, we can assume that $\tB$ varies little over the finite radial and vertical extent of the detector's FOV. Thus, we expect that horizontally-polarized light is \emph{only} measurable for $\tp \in [\tpmin,\tpmax]$. This implies that the $\pot$ transition in $\tpol$ measurements (horizontal to vertical) should be observed as a RE's pitch angle increases past $\tp \approx \tpmax$. Additionally, for REs with $\tp \approx \tpmax$, we hypothesize that a mix of light with many polarizations will lead to low values of $\fpol$. Finally, we expect that a detector should see almost no synchrotron radiation from a RE with pitch angle $\tp < \tpmin$ because the RE is simply not emitting light into the detector's FOV. 

The radial profile of $\tp \in [\tpmin, \tpmax]$ for a sample magnetic geometry from EFIT~\cite{lao1985} is shown as the bounded region in figure~\ref{fig:paRef}; \sout{these are the} \red{this is the range of RE} pitch angles within which a \red{synchrotron polarization} measurement of $\tpol \approx 0\deg$ is expected. As will be discussed further in section~\ref{sec:response}, this simple model shows good agreement when compared to synthetic data from \SOFT. Notably, measurements of $\tpol \approx 90\deg$ are more probable close to the magnetic axis, and the minimum RE pitch angle at which synchrotron emission is detected increases with increasing $\rtan/a$.  


\red{The previous} \sout{These} and following calculations require knowledge of the magnetic field geometry. For RE populations carrying a significant fraction of the plasma current, like RE plateaus after disruptions in some tokamaks \cite{plyusnin2006,martin-solis2006,plyusnin2018}, the interpretation of synchrotron polarization data, specifically, would become more difficult and convoluted. However, in C-Mod, almost no variation in the externally-applied loop voltage is observed during flattop RE discharges as synchrotron emission increases in time; therefore, the RE current is inferred to be negligible compared to the total plasma current, and a magnetic reconstruction, like EFIT, is considered to be an accurate approximation of the real magnetic topology. \red{Note that due to observations of the sawtooth instability in the electron temperature, the safety factor at the magnetic axis was constrained to be in the range $q_{\rm axis} \in 0.95 \pm 0.3$, which is standard for EFIT reconstructions of C-Mod plasmas. Nevertheless, the final results were found to be relatively insensitive to this bound.}


\subsection{A numerical model for polarized synchrotron radiation}\label{sec:SOFT}

The numerical tool \SOFT\ uses Stokes parameters, as introduced in section~\ref{sec:synchrotron}, to represent the polarization of synchrotron radiation. Since the purpose of \SOFT\ is to simulate the signals reported by synchrotron radiation diagnostics, the definition of the Stokes parameters used must correspond to those used by the diagnostic. For example, the definition in section~\ref{sec:synchrotron} uses a radiation-local coordinate system, whereas a diagnostic will measure the radiation in a fixed coordinate system that is independent of the propagation direction of the radiation.

\begin{figure}[h!]
    \centering
    \includegraphics{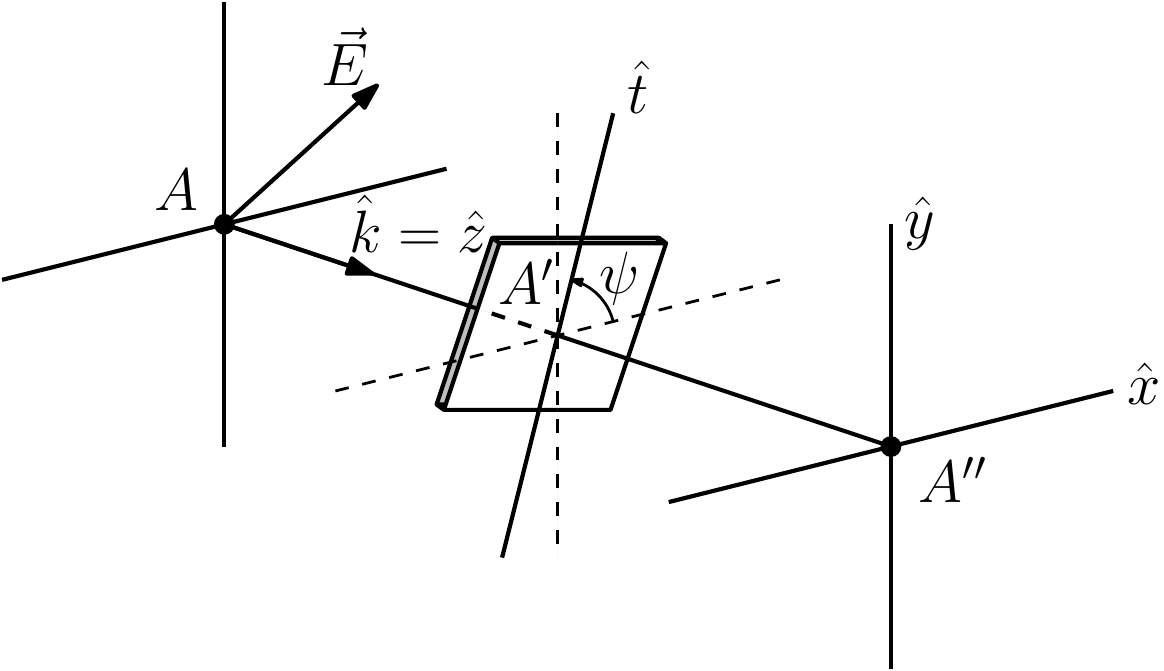}
    \caption{Illustration of an idealized setup for measuring the four Stokes parameters. Polarized radiation emitted at $A$ along $\hat{k}=\hat{z}$ is incident on the linear polarization filter with transmission axis $\hat{t}$ at $A'$. By rotating the polarization filter between $\psi = 0, \pi/4$, and $\pi/2$, we can solve for $I$, $Q$, and $U$ from the measured intensity at $A''$. To also obtain $V$, we introduce a phase-shift between the electric field components by putting a quarter-wave plate in front of the linear polarizer.}
    \label{fig:stokes_setup}
\end{figure}

A simple, yet relatively general model for a polarization-measuring diagnostic can be obtained starting from the idealized setup shown in figure~\ref{fig:stokes_setup}, where a polarizer has been placed between the emitter at $A$ and the observer at $A''$. It can be shown~\cite{stone1963} that if the polarizer consists of just a linear polarization filter, with its transmission axis $\hat{t}$ rotated about the $\hat{z}$ axis by an angle $\psi$ from the horizontal, then the irradiance at $A''$ is given by
\begin{equation}\label{eq:irrLin}
    \mathcal{I}(\psi) = \frac{1}{2}\left( I + Q\cos 2\psi + U\sin 2\psi \right).
\end{equation}
If we were to put a quarter-wave plate in front of the linear polarizer, so that the relative phase of the electric field components is shifted, the irradiance measured at $A''$ \red{would be} \sout{is}
\begin{equation}\label{eq:irrCirc}
    \mathcal{I}_{\lambda/4}(\psi) = \frac{1}{2}\left( I + Q\cos 2\psi + V\sin 2\psi \right).
\end{equation}
By measuring the irradiance at $A''$ with the linear polarizer rotated to $\psi = 0, \pi/4$, and $\pi/2$, as well as at $\psi = \pi/4$ with the quarter-wave plate, we can solve for the Stokes parameters and find
\begin{eqnarray}
    I &= \mathcal{I}(0) + \mathcal{I}(\pi/2),\label{eq:StokesParamI}\\
    Q &= \mathcal{I}(0) - \mathcal{I}(\pi/2),\\
    U &= 2\mathcal{I}(\pi/4) - \mathcal{I}(0) - \mathcal{I}(\pi/2),\\
    V &= 2\mathcal{I}_{\lambda/4}(\pi/4) - \mathcal{I}(0) - \mathcal{I}(\pi/2).\label{eq:StokesParamV}
\end{eqnarray}
In what follows, we shall take~\eqref{eq:StokesParamI}-\eqref{eq:StokesParamV} as definitions of the Stokes parameters. In the idealized setup of figure~\ref{fig:stokes_setup}, where radiation is incident on the polarizer perpendicularly, these definitions correspond to the usual definitions in section~\ref{sec:synchrotron}. However, when radiation is incident on the polarizer along $\hat{k}$ and makes an angle with $\hat{z}$, the relations~\eqref{eq:StokesParamI}-\eqref{eq:StokesParamV} become approximate, with errors of order $1-(\hat{k}\cdot\hat{z})^2$. Since the angle between $\hat{k}$ and $\hat{z}$ is of the order of the FOV opening angle $\alpha$, the error is less than $10^{-4}$ in the present setup, making \eqref{eq:StokesParamI}-\eqref{eq:StokesParamV} a good approximation of the measured Stokes parameters.

In \SOFT, the effect of the polarizer enters through the power per solid angle $\Omega$ and wavelength $\lambda$, generally given by $\dd^2 P(\psi) / \dd\lambda\dd\Omega = R^2\mathcal{I}$, measured by a detector located behind the polarizer at a distance $R$ from the emitting electron. Since electrons may be located anywhere in the tokamak, the direction of propagation $\vec{k}$ of the radiation and polarizer surface normal $\hat{z}$ are usually not parallel; this must be taken into account when modeling the linear polarizer. In general, the irradiance is equal to
\begin{equation}
    \mathcal{I}(\psi) = \epsilon_0 c \left| \vec{E}_{\rm in} \right|^2 = \epsilon_0 c \left| T\vec{E} \right|^2,
\end{equation}
where $\vec{E}_{\rm in}$ is the incident electric field vector, the operator $T$ describes the action of the polarizer, $\epsilon_0$ denotes the permittivity of free space, and $c$ the speed of light in vacuum. In the case when the polarizer consists of both a linear polarization filter and a quarter-wave plate, we can write $T$ as the product between two matrices describing the action of each element, i.e. $T = T_{\rm p} T_{\lambda/4}$, where $T_{\rm p}$ describes the action of the linear polarizer and $T_{\lambda/4}$ the action of the quarter-wave plate. If $T_{\lambda/4}$ is oriented with fast axis along the detector $\hat{x}$-axis (and thus slow axis along $\hat{y}$), $T_{\lambda/4}$ can be written in dyadic notation as
\begin{equation}
    T_{\lambda/4} = \hat{x}\hat{x} + e^{-i\pi/2}\hat{y}\hat{y} + \hat{z}\hat{z}.
\end{equation}
For the linear polarization filter, represented by $T_{\rm p}$, we can use the model presented in~\cite{korger2013}, which describes the action of a linear polarization filter on obliquely incident radiation. The model assumes that the filter absorbs radiation along its absorption axis $\hat{a}$, perpendicular to the transmission axis, so that in the setup of figure~\ref{fig:stokes_setup} the absorption axis becomes $\hat{a} = \hat{y}\cos\psi - \hat{x}\sin\psi$. When the radiation is not incident on the filter perpendicularly as in figure~\ref{fig:stokes_setup}, i.e.\ when $\hat{k}\neq\hat{z}$, radiation is instead assumed to be absorbed along an \emph{effective absorption axis} $\hat{a}_{\rm eff}$, which is the projection of $\hat{a}$ on the plane of the polarized radiation, normalized: 
\begin{equation}
    \hat{a}_{\rm eff} = \frac{\hat{a} - \hat{k}\left( \hat{a}\cdot\hat{k}\right)}{\sqrt{1 - \left( \hat{a}\cdot\hat{k} \right)^2}}.
\end{equation}
Using $\hat{a}_{\rm eff}$, the matrix $T_{\rm p}$ for the linear polarizer can be written
\begin{equation}
    T_{\rm p} = \I - \hat{a}_{\rm eff}\hat{a}_{\rm eff},
\end{equation}
where $\I$ denotes the identity matrix. Hence, the measured irradiance is
\begin{eqnarray}
    \mathcal{I}(\psi) &=  \epsilon_0 c\left| \vec{E} - \hat{a}_{\rm eff}\left( \hat{a}_{\rm eff}\cdot\vec{E} \right) \right|^2,\\
    \mathcal{I}_{\lambda/4}(\psi) &=  \epsilon_0 c\left| T_{\lambda/4}\vec{E} - \hat{a}_{\rm eff}\left[ \hat{a}_{\rm eff} \cdot\left( T_{\lambda/4} \vec{E} \right) \right] \right|^2,
\end{eqnarray}
without and with the quarter-wave plate, respectively.

    %
    %
    %


\subsection{Detector response functions}\label{sec:response}

One powerful feature of \SOFT is its calculation of Green's functions $\Ghat(R,p,\tp)$ for quantities of interest such as the Stokes parameters. These are essentially detector response functions which, when integrated with a RE phase space distribution, give the expected synthetic signal. These response functions account for the detector geometry and spectral range, as well as the magnetic field. The functional dependencies of $\Ghat$ on \red{the} radius $R$, total momentum $p$, and pitch angle $\tp$ allow us to better constrain the RE population in position and momentum space when comparing synthetic and experimental measurements. For this work, a response function was computed for each channel (and each time of interest) using parameters in table~\ref{tab:MSE}, magnetic geometries from EFIT, and a phase space spanning $r/a \in [0,1]$, $p/mc \in [0,100]$ (i.e. energies extending up to approximately 50~MeV), and $\tp \in [0,0.3]$~rad. 

\begin{figure}[h!]
    \centering
    \includegraphics[width=\myWidth]{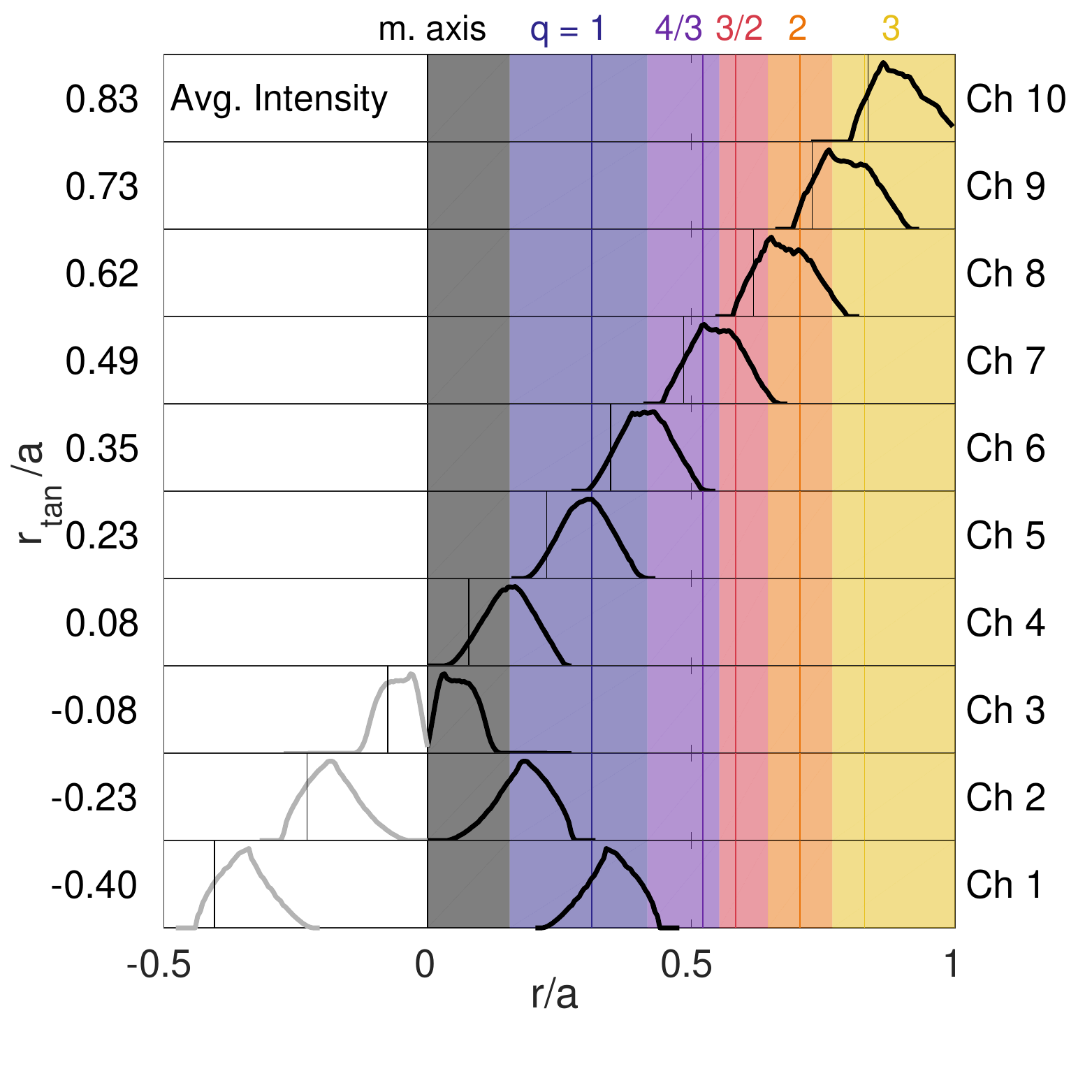}
    \caption{Synthetic intensity measurements of each MSE channel modeled in \SOFT, averaged with a uniform distribution over momentum space, as a function of normalized minor radius (i.e. initial RE position). The experimentally-determined normalized tangency radius is indicated by a vertical black line for each channel (labeled at left). Note that channels 1-3 have data reflected over $r/a = 0$ (in grey). The locations of the magnetic axis \red{(with $q_{\rm axis} \approx 0.9$)} and flux surfaces $q$~=~1, 4/3, 3/2, 2, and 3 are shown as solid vertical lines; \redd{shaded regions, extending halfway between adjacent surfaces, are used in step~\ref{step:stitch} of the methodology of section~\ref{sec:comparison}.}} 
    \label{fig:IvR}
\end{figure}

Figure~\ref{fig:IvR} shows the intensity $\langle\hat{I}\rangle$ of light \emph{detected} by each channel (and averaged over momentum space) versus radius for REs populated across the entire plasma. \red{As expected,} \sout{Note how} each channel's measurements are radially-localized,\sout{ as expected,} approximately at the appropriate $\rtan/a$, indicated as a black vertical line for each channel. In \SOFT, REs are ``initiated'' on the outer midplane ($R \geq R_0$) and then follow magnetic field lines. Therefore, while some channels only ``see'' REs on the high field side ($R < R_0$), the Green's function records their starting positions on the low field side. This is the case for channels~1-3 (see figure~\ref{fig:mse}), so their \SOFT-measured synthetic intensities have been mirrored over the magnetic axis in figure~\ref{fig:IvR}, as shown in grey. There is significant overlap in radial distributions of detected intensity between adjacent channels, and some pairs of channels even view almost the same radial band of REs. For instance, channels~1 and 2 are sensitive to the same REs as channels~5 and 4, respectively. Thus, an opportunity exists to interrogate different parts of the \emph{same} RE phase space distribution by comparing data between multiple channels. 

\begin{figure}[h!]
    \centering
    \begin{subfigure}{\halfWidth}
        \centering
        \includegraphics[width=\textwidth]{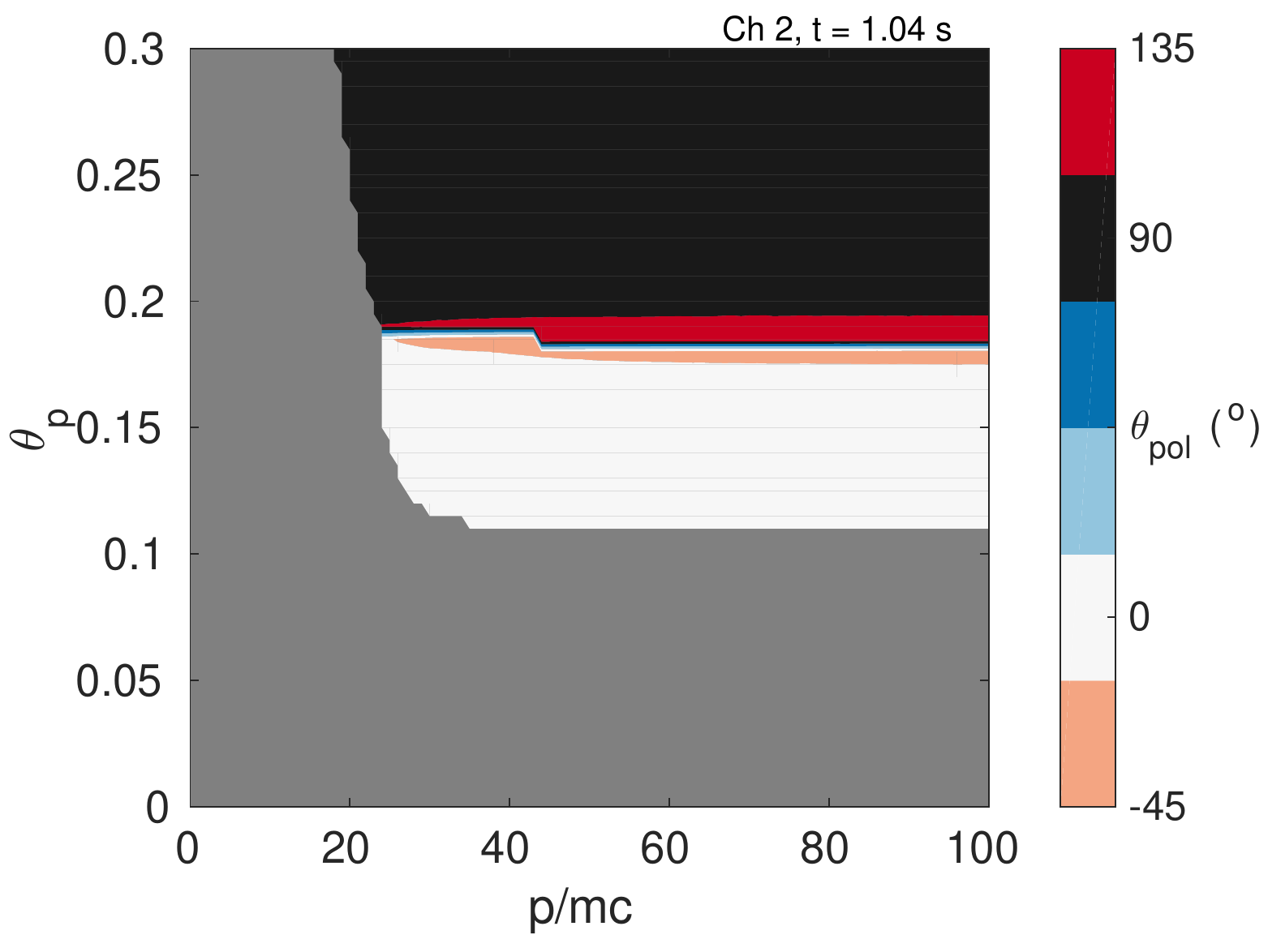}
        \caption{}
        \label{fig:pa_ch2}
    \end{subfigure}
    \begin{subfigure}{\halfWidth}
        \centering
        \includegraphics[width=\textwidth]{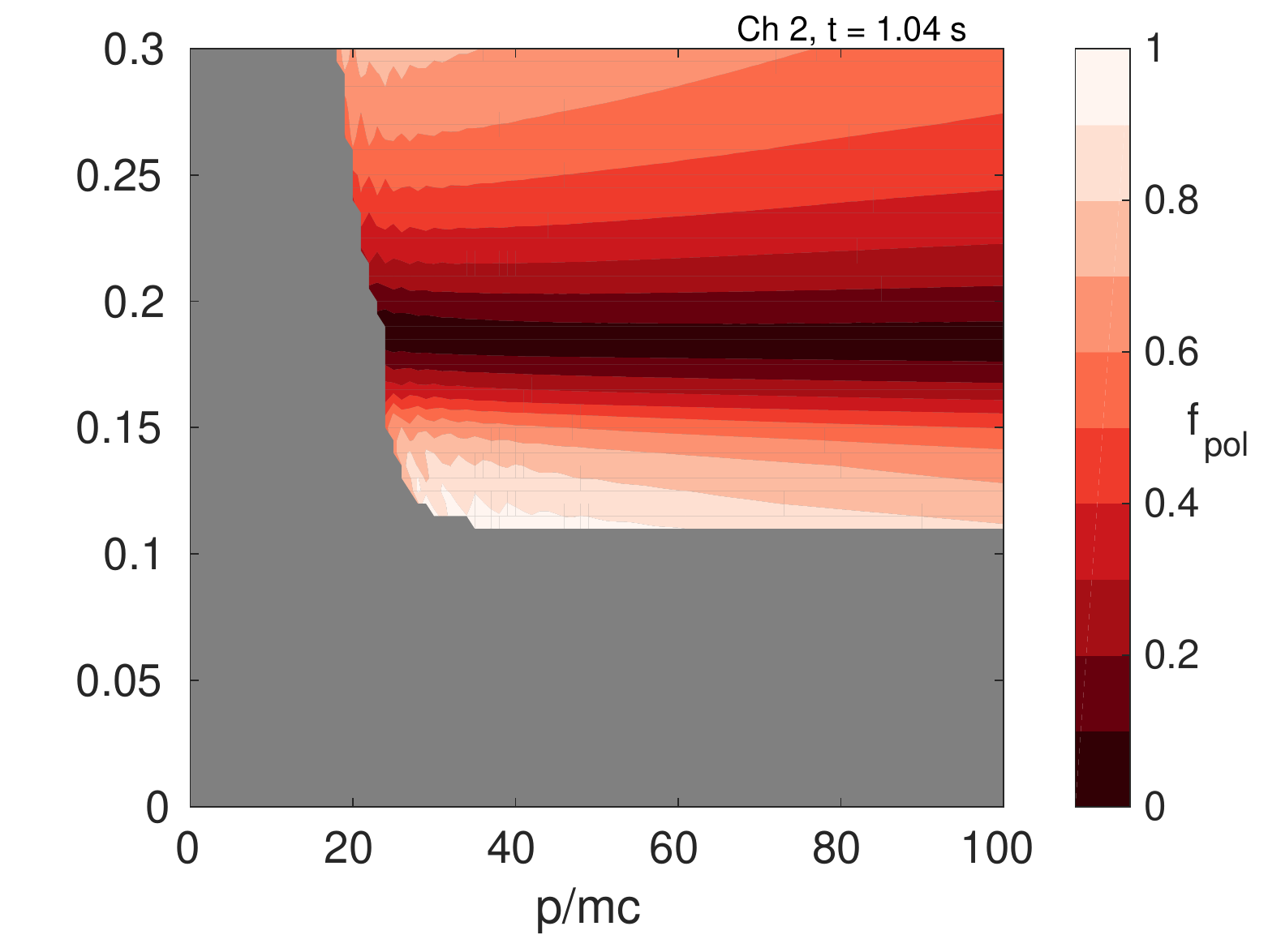}
        \caption{}
        \label{fig:pf_ch2}
    \end{subfigure}
    \caption{Detector response functions, from \SOFT, of the polarization (a)~angle $\tpol$ and (b)~fraction $\fpol$ versus normalized (total) momentum $p/mc$ and pitch angle $\tp$, for MSE channel~2 at time $t$~=~1.04~s. Grey regions indicate practically-undetectable regions of phase space.}
\end{figure}

Sample response functions of $\tpolhat$ and $\fpolhat$ are plotted over momentum space for channel~2 ($\rtan/a = -0.23$) in figures~\ref{fig:pa_ch2} and \ref{fig:pf_ch2}, respectively. These have already been integrated over position space using the RE density profile inferred from \red{visible/near-infrared} images of synchrotron emission \cite{tinguely2018ppcf}; in any case, the results are fairly insensitive to the density profile shape. The interpretation of these response functions is that a single RE with a given momentum and pitch---i.e. a delta function in momentum space---would produce synchrotron emission resulting in the shown measurement of $\tpol$ or $\fpol$. The grey areas in both plots indicate the regions in which little-to-no signal\footnote{Specifically, the cutoff for \SOFT data was (arbitrarily) chosen to be $L/\max(L) \leq 10^{-8}$ for all channels and times. The final results are insensitive to this choice.} is detected by channel~2. Therefore, this specific channel geometry limits the diagnosis of REs to those with momenta $p/mc > 20$ and pitch angles $\tp > 0.11$~rad. It is clear from figure~\ref{fig:pa_ch2} that measurements of $\tpol = 0\deg$ or $90\deg$ are most common, and the phase space is bifurcated at a critical pitch angle $\tpc \approx 0.185$~rad. This implies that if channel~2 records a measurement of $\tpol \approx 90\deg$, then a significant fraction of the RE population must have pitch angles $\tp > \tpc$. Conversely, measuring $\tpol \approx 0\deg$ implies that the bulk of the RE distribution function is confined within $\tp < \tpc$. An improved localization of REs in momentum space can be obtained by using both $\tpolhat$ and $\fpolhat$ data. In figure~\ref{fig:pf_ch2}, the minimum of $\fpolhat$ occurs near $\tpc$ and increases as $\vert\tp - \tpc\vert$ increases, as expected from the heuristic argument presented in section~\ref{sec:heuristic}. Thus, for example, experimental measurements of $\tpol \approx 90\deg$ and $\fpol \approx 0.5$ would indicate that the detected synchrotron light is dominated by REs with pitch angles $\tp \approx 0.25$~rad.

\begin{figure}[h!]
    \centering
    \begin{subfigure}{\halfWidth}
        \centering
        \includegraphics[width=\textwidth]{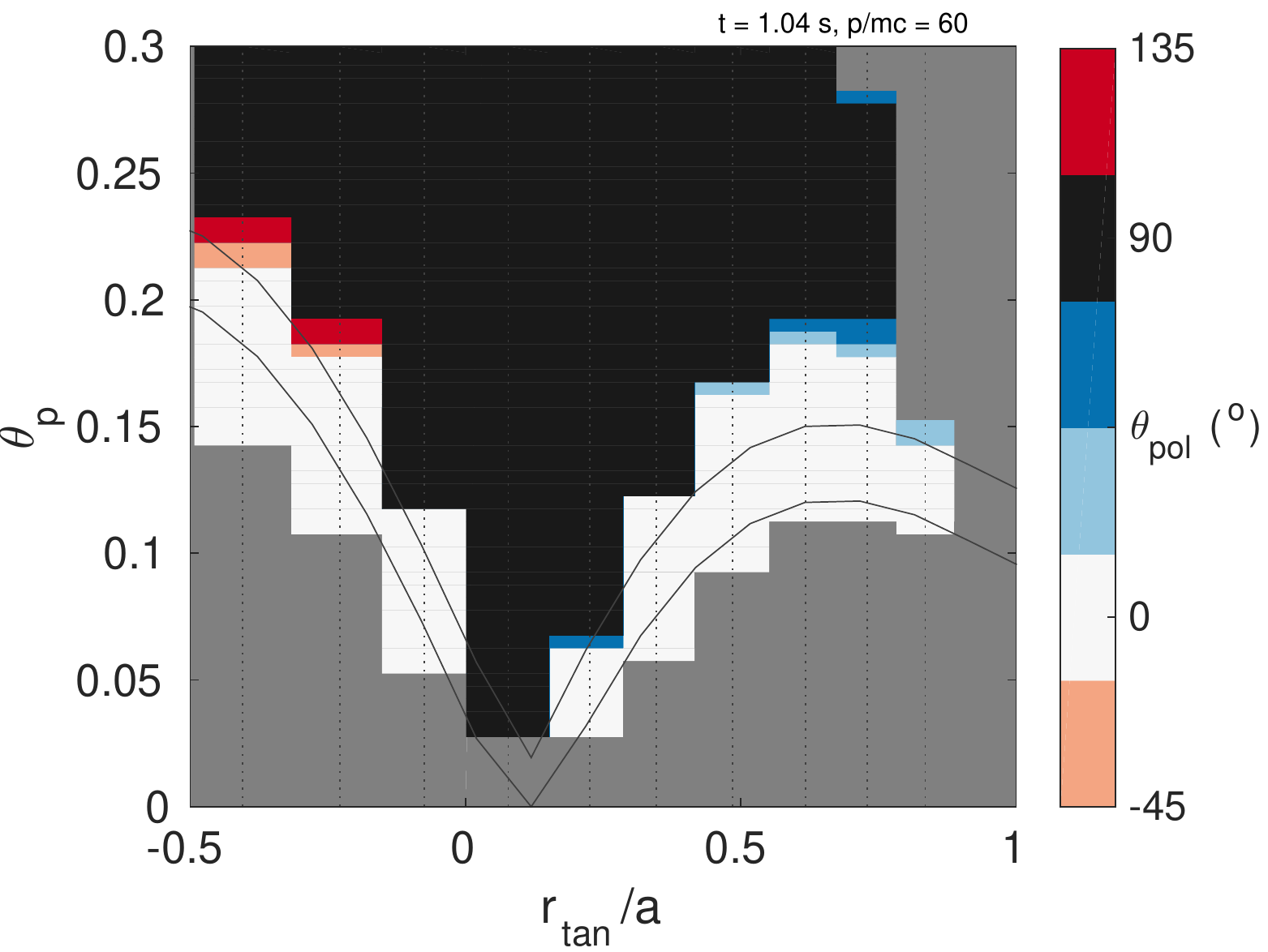}
        \caption{}
        \label{fig:paRef}
    \end{subfigure}
    \begin{subfigure}{\halfWidth}
        \centering
        \includegraphics[width=\textwidth]{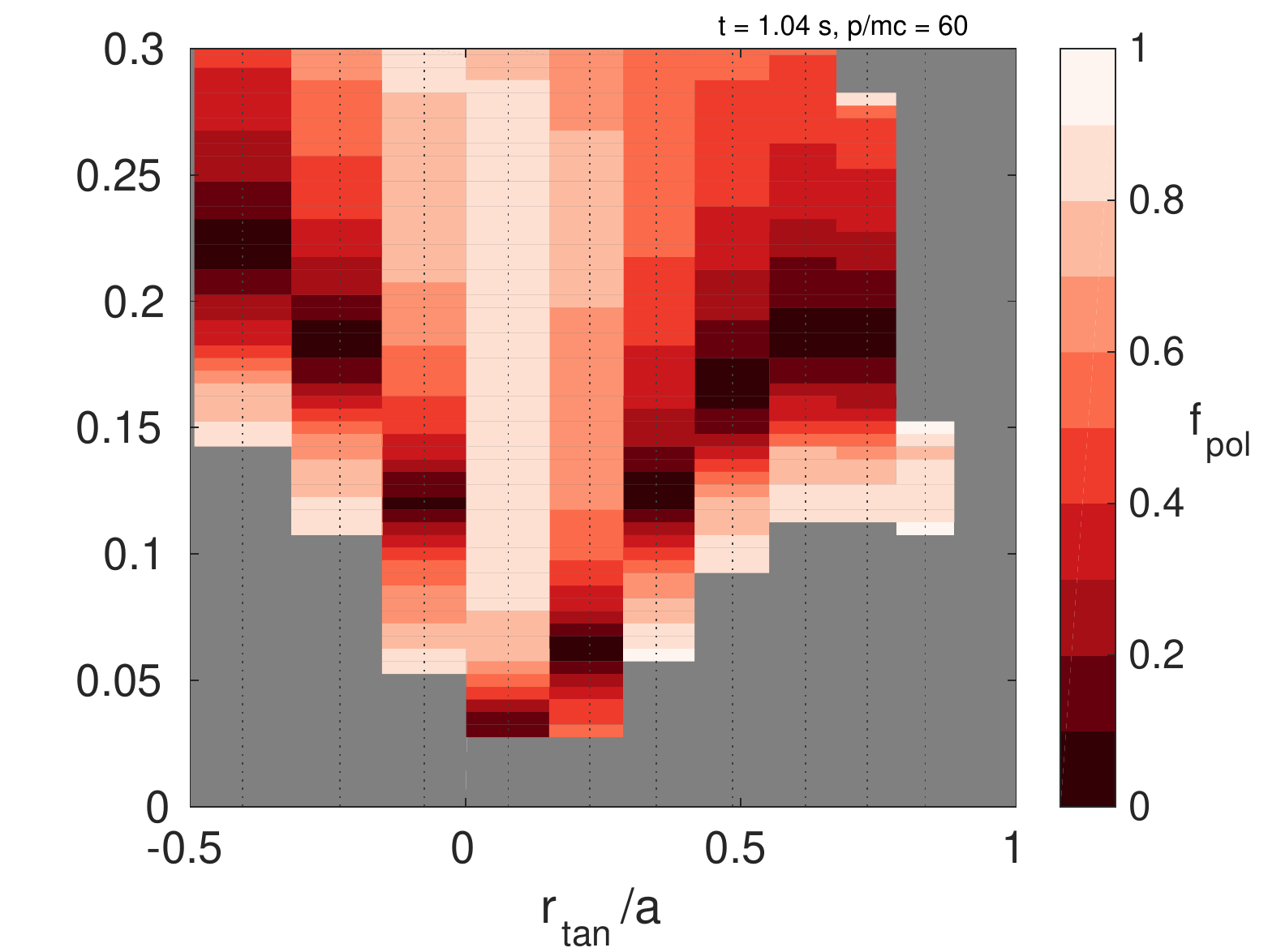}
        \caption{}
        \label{fig:pfRef}
    \end{subfigure}
    \caption{\SOFT-predicted polarization (a)~angle $\tpol$ and (b)~fraction $\fpol$ versus normalized tangency radius $\rtan/a$ of the MSE channels (vertical dotted lines) and RE pitch angle $\tp$, for $t$~=~1.04~s and $p/mc$~=~60. The bounded region in (a) corresponds to the region of expected $\tpol \approx 0\deg$ from the heuristic argument presented in section~\ref{sec:heuristic}, i.e. $\tpmin \leq \tp \leq \tpmax$ \red{from \eqref{eq:heuristicCalc}}. See figure~\ref{fig:linePlots} for line plots of $\tpol$ and $\fpol$ for channels~1, 3, 5, 7, and 9. \red{Grey regions indicate practically-undetectable regions of phase space.}}
    \label{fig:Ref}
\end{figure}

\begin{figure}[h!]
    \centering
    \includegraphics[width=\myWidth]{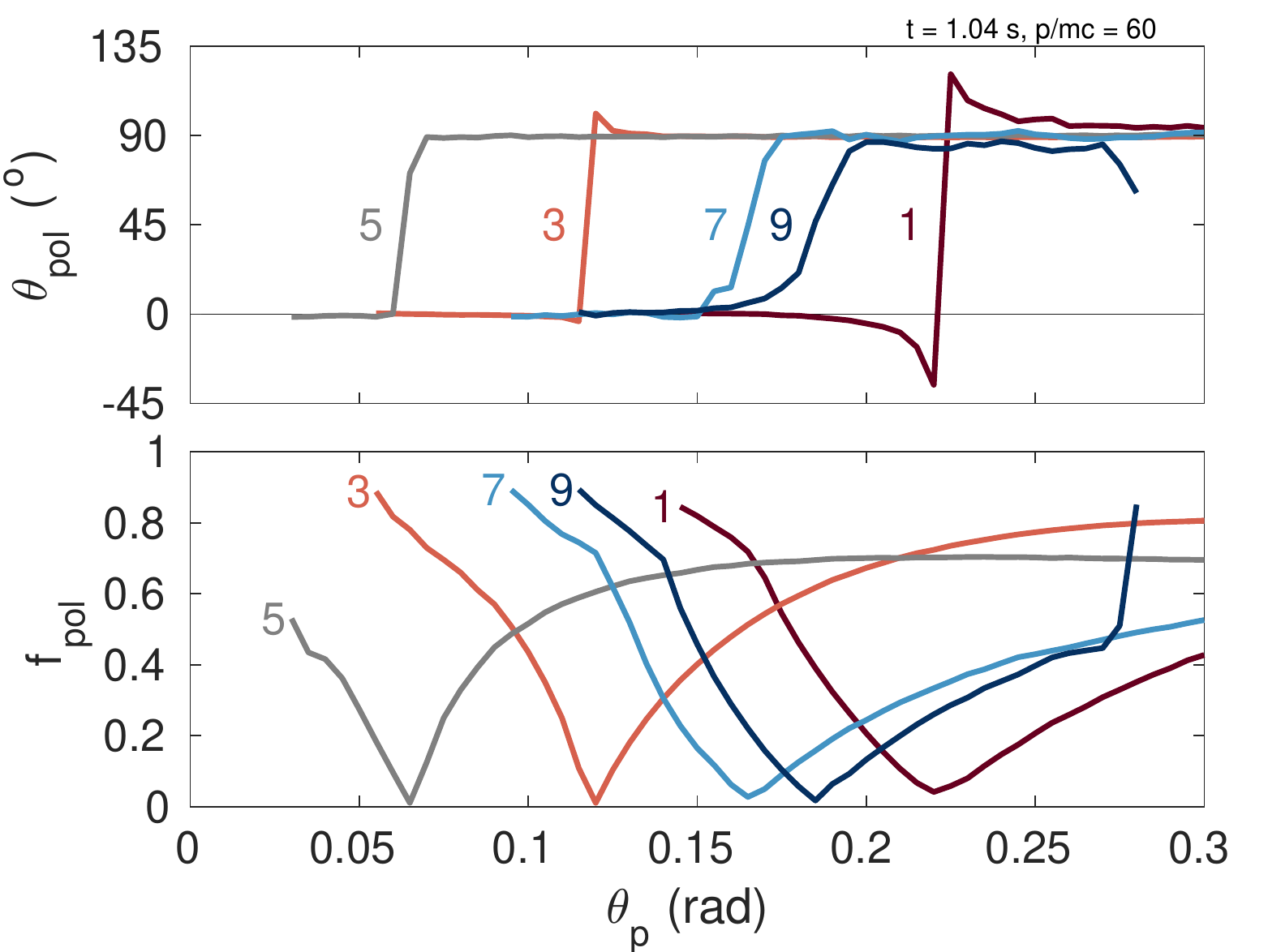}
    \caption{Trends in \SOFT-predicted polarization (top) angle $\tpol$ and (bottom) fraction $\fpol$ versus RE pitch angle $\tp$ for channels~1, 3, 5, 7, and 9 (labeled). The data for these line plots correspond to the \emph{same} data in the contour plots of figures~\ref{fig:paRef} and \ref{fig:pfRef}, re-plotted here for clarity.}
    \label{fig:linePlots}
\end{figure}

As seen in figures~\ref{fig:pa_ch2} and \ref{fig:pf_ch2}, synthetic measurements are relatively insensitive to the RE momentum $p/mc$. As mentioned, they are also insensitive to the RE density profile $\nRE$ (not shown) for two main reasons: (i) measurements are dominated by REs within narrow radial bands (see figure~\ref{fig:IvR}) over which significant variations in $\nRE$ are not expected, and (ii) both $\tpol$ and $\fpol$ are independent of the emission amplitude. Therefore, a cross-section of $\Ghat(R,p,\tp)$, at one momentum and summed over each channel's radial range, provides a reference or ``look-up'' plot of a synchrotron polarization measurement versus channel $\rtan/a$ and RE pitch angle $\tp$. These are shown for $\tpol$ and $\fpol$ in the contour plots in figures~\ref{fig:paRef} and \ref{fig:pfRef}, respectively, as well as the line plots in figure~\ref{fig:linePlots}.

In figure~\ref{fig:paRef}, note that only a narrow band of $\tpol \approx 0\deg$ (white) is expected; it follows a similar spatial pattern as that predicted by the heuristic model in section~\ref{sec:heuristic}, with bounds of $\tpmin$ and $\tpmax$ \red{from \eqref{eq:heuristicCalc}} overlaying the data. The differences between \red{the} heuristic argument and simulation here are likely due to the more realistic 3D detector geometry implemented in \SOFT. \red{In addition, recall that these calculations were performed using the magnetic geometry from the ``standard'' EFIT reconstruction, as described previously; the uncertainty in the bounds of the region $\tpol \approx 0\deg$, resulting from EFIT reconstruction error, is expected to be $\vert\Delta\tp\vert \lesssim 0.02$~rad.} The complementary reference plot for $\fpol$ is shown in figure~\ref{fig:pfRef}, where it is again seen that the minimum in $\fpol$ always occurs at the $\pot$ transition location seen in figure~\ref{fig:paRef}. The same data in figures~\ref{fig:paRef} and \ref{fig:pfRef} are also shown in figure~\ref{fig:linePlots} for only channels~1, 3, 5, 7, and 9; these plots are \red{provided} \sout{simply} to help the reader better visualize the spatial (between-channel) variations of the $\pot$ transition in $\tpol$ and minimum in $\fpol$.


Comparing these plots to C-Mod experimental data can help identify the pitch angles of REs which dominate the synchrotron emission measurement. However, not all channels provide useful information. For example, channel~4 ($\rtan/a = 0.08$) should \emph{always} measure a polarization angle of $\tpol \approx 90\deg$, which is confirmed by experiment as seen in figure~\ref{fig:hist_pa}. In addition, the $\fpol$ measurement from channel~4 should be maximal for typical RE pitch angles $\tp > 0.05$~rad; this is also seen in the experimental trends of figure~\ref{fig:hist_pf}. \redd{Values of $\fpol \leq 0.6$ could help constrain the pitch angle to \red{$\tp \approx 0.05$~rad}; however, it is difficult to make a quantitative comparison in this situation since background/reflected light is not yet included in \SOFT.} Finally, note that data in figures~\ref{fig:paRef} and \ref{fig:pfRef} are not perfectly symmetric about the magnetic axis; therefore, the seemingly conflicting measurements between channels which ``see'' REs on the same flux surface (approximately $\pm \rtan/a$) are actually the result of each channel investigating a different region of phase space.


\section{Comparisons of experimental and synthetic data}\label{sec:comparison}

\begin{figure}[h!]
    \centering
    \begin{subfigure}{\halfWidth}
        \centering
        \includegraphics[width=\textwidth]{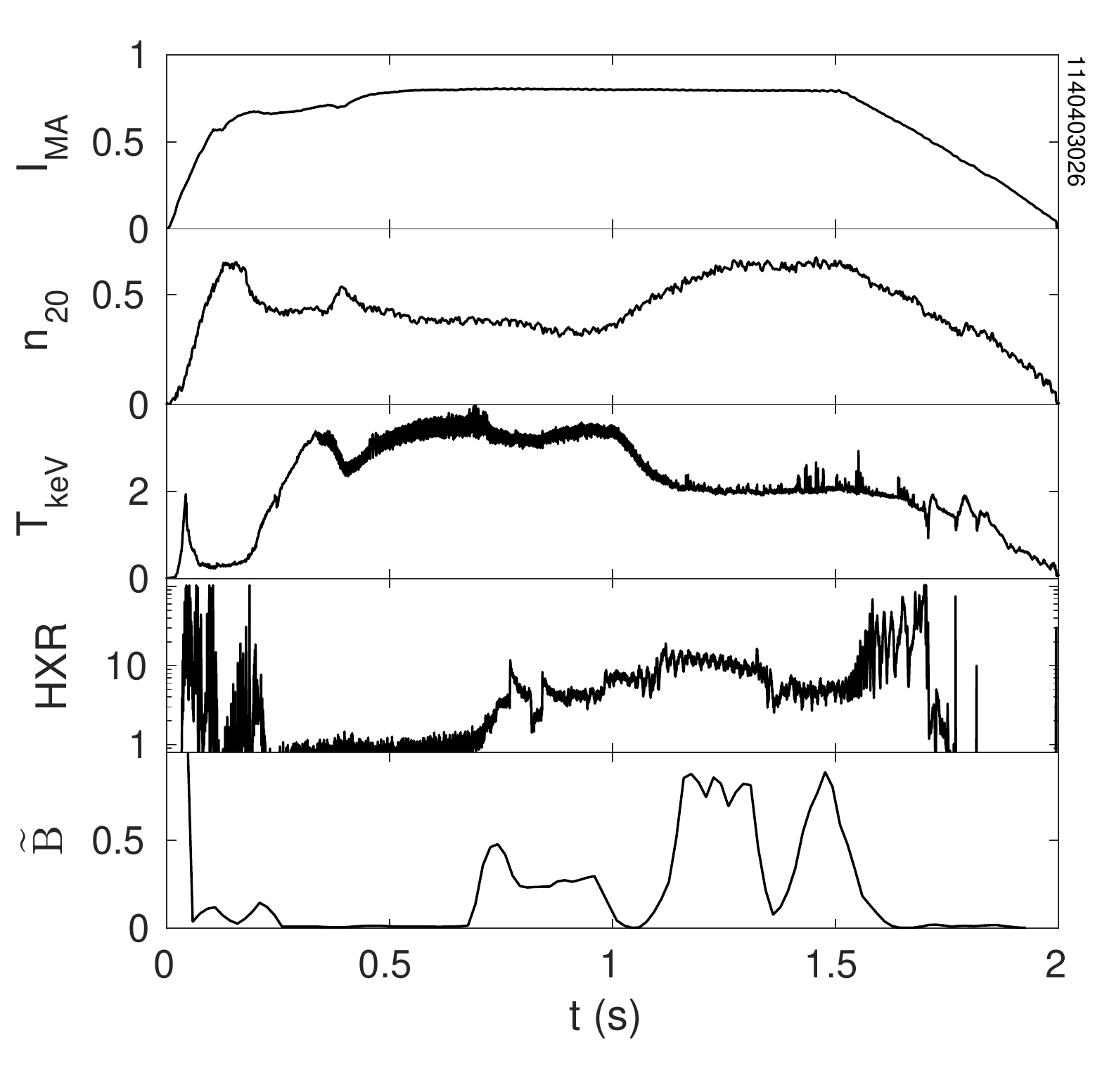}
        \caption{}
        \label{fig:params}
    \end{subfigure}
    \begin{subfigure}{\halfWidth}
        \centering
        \includegraphics[width=\textwidth]{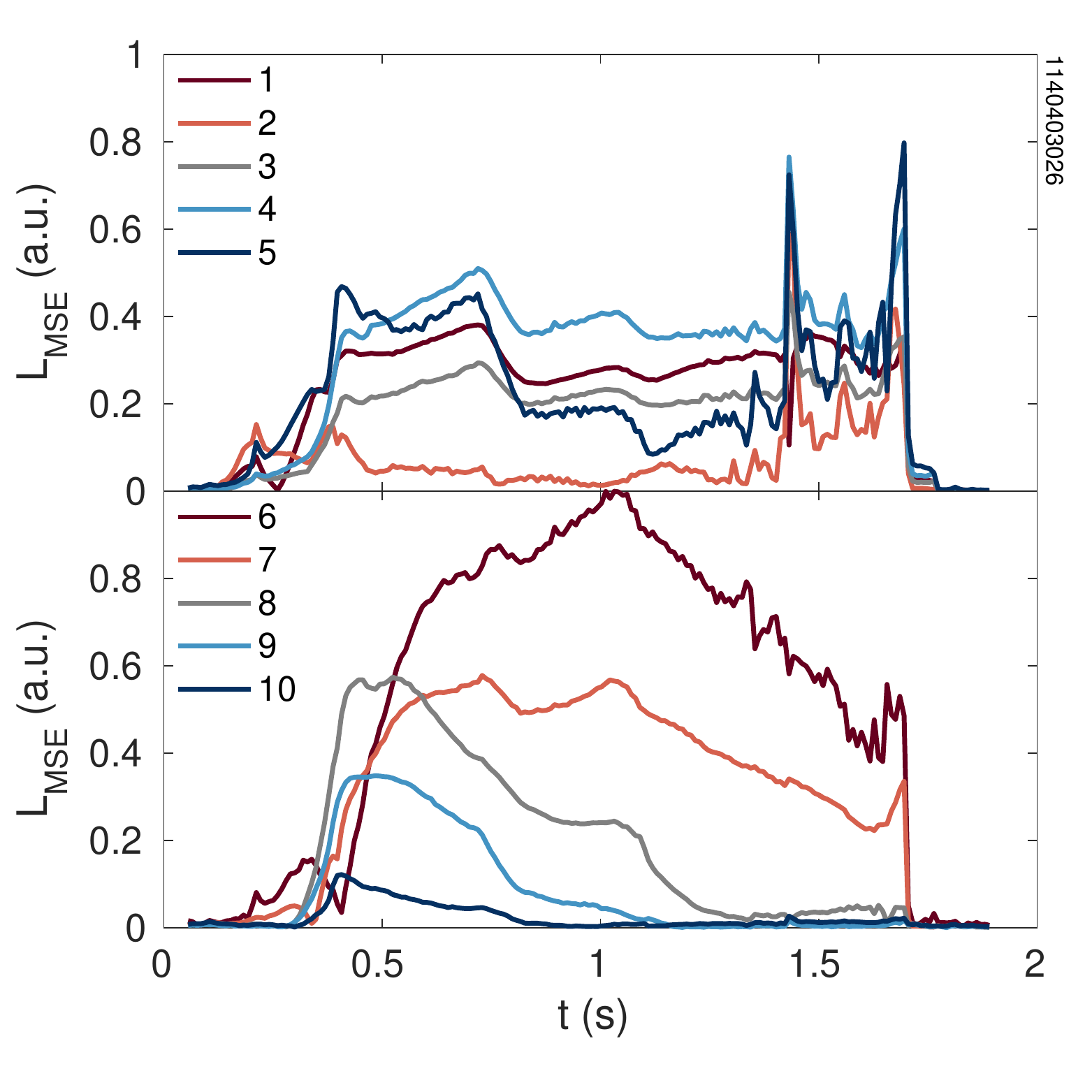}
        \caption{}
        \label{fig:LMSE}
    \end{subfigure}
    \caption{(a) From top to bottom: the plasma current (MA), line-averaged electron density ($10^{20}$~m$^{-3}$), central electron temperature (keV), HXR signal (a.u.), and locked mode indicator (a.u.) are shown for an Alcator C-Mod plasma discharge. (b) The intensity of linearly-polarized light (a.u.) from MSE channels~1-5 (top) and 6-10 (bottom), all normalized to the maximum of channel~6.}
    \label{fig:paramsAndL}
\end{figure}

In this section, the spatiotemporal evolution of polarized synchrotron emission is explored in detail for one Alcator C-Mod discharge. Note that this is the same discharge for which images of synchrotron light were analyzed in \red{the visible/near-infrared wavelength range, $\lambda \approx$~400-900~nm} \cite{tinguely2018ppcf}. Time traces of several plasma parameters are shown in figure~\ref{fig:params}. In this experiment, REs were generated during the flattop portion of the plasma current $\IMA$ by decreasing the plasma density $n_{20}$ and hence collisional friction. The intensity of linearly-polarized synchrotron emission $\LMSE$, shown for all channels in figure~\ref{fig:LMSE}, starts rising at $t \approx$~0.4~s. Recall that there is not a reliable absolute calibration of $\LMSE$ among the channels; relative errors of 30-40\% are expected. However, there are certainly similar temporal trends, especially in groups of channels~1-5, 6-7, and 8-10.

The hard x-ray (HXR) signal increases at $t \approx$~0.7~s; this is around the same time that a locked mode begins, indicated by a reduction of plasma temperature $\TkeV$ sawteeth and an increase in magnetic fluctuations $\widetilde{B}$. This likely indicates the expulsion of REs from the plasma due to MHD activity, leading to thick-target bremsstrahlung with the first wall. Dips in the intensity $\LMSE$ are also observed across all channels at this time. 

At $t$~=~1~s, the density is increased to suppress REs, and $\LMSE$ starts to decrease, in particular for channels~6-10; these are viewing the RE beam edge as it shrinks in size. Sharp spikes in both HXR and $\LMSE$ signals---especially across channels~1-5---begin at $t\approx$~1.5~s, around the time of the final burst of MHD activity and ramp-down in plasma current and density. Both the synchrotron emission and HXR signals then disappear at $t \approx$~1.7~s.

\begin{figure}[h!]
    \centering
    \begin{subfigure}[h]{\halfWidth}
        \centering
        \includegraphics[width=\textwidth]{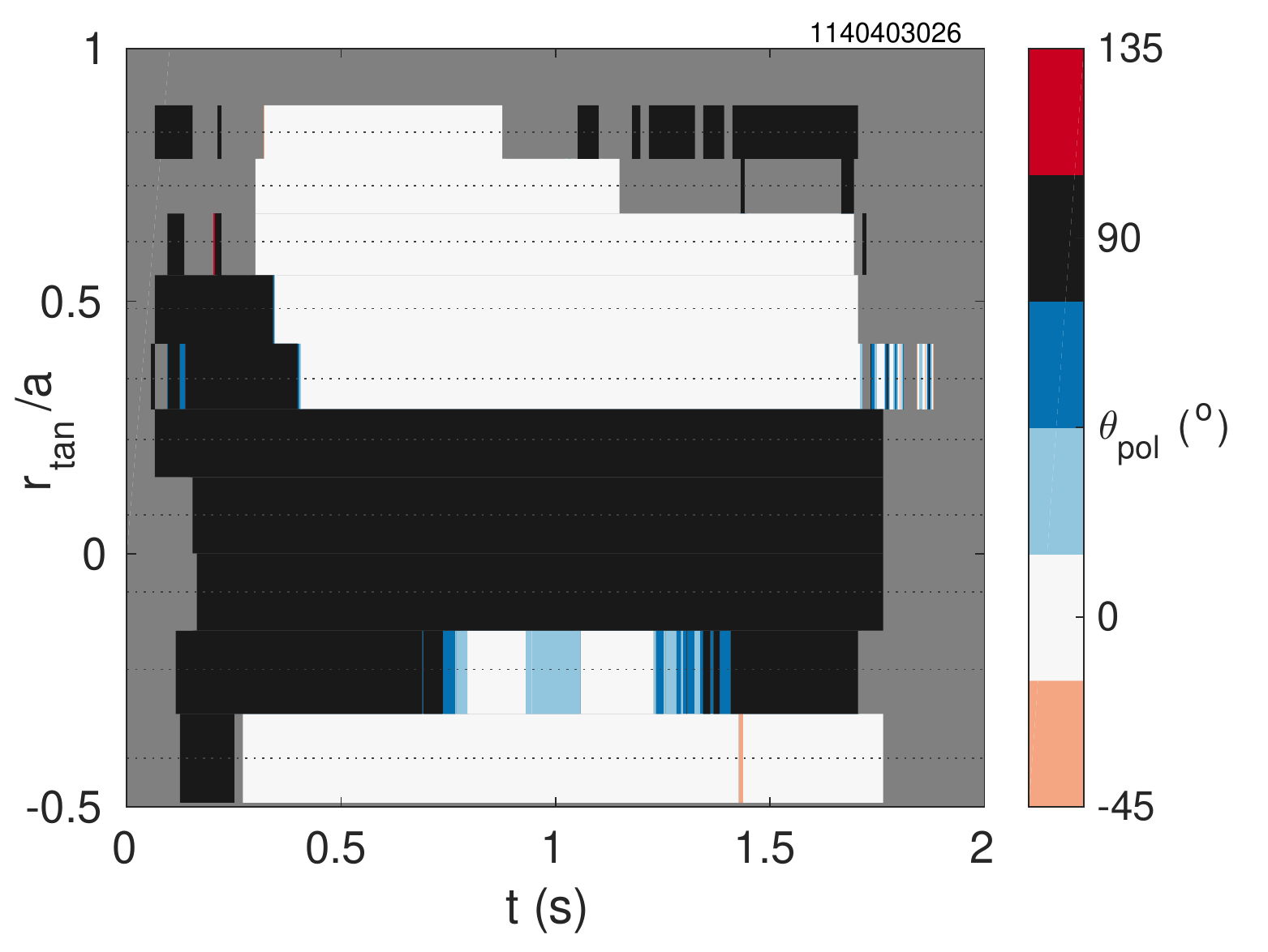}
        \caption{}
        \label{fig:paExp}
    \end{subfigure}
    \begin{subfigure}{\halfWidth}
        \centering
        \includegraphics[width=\textwidth]{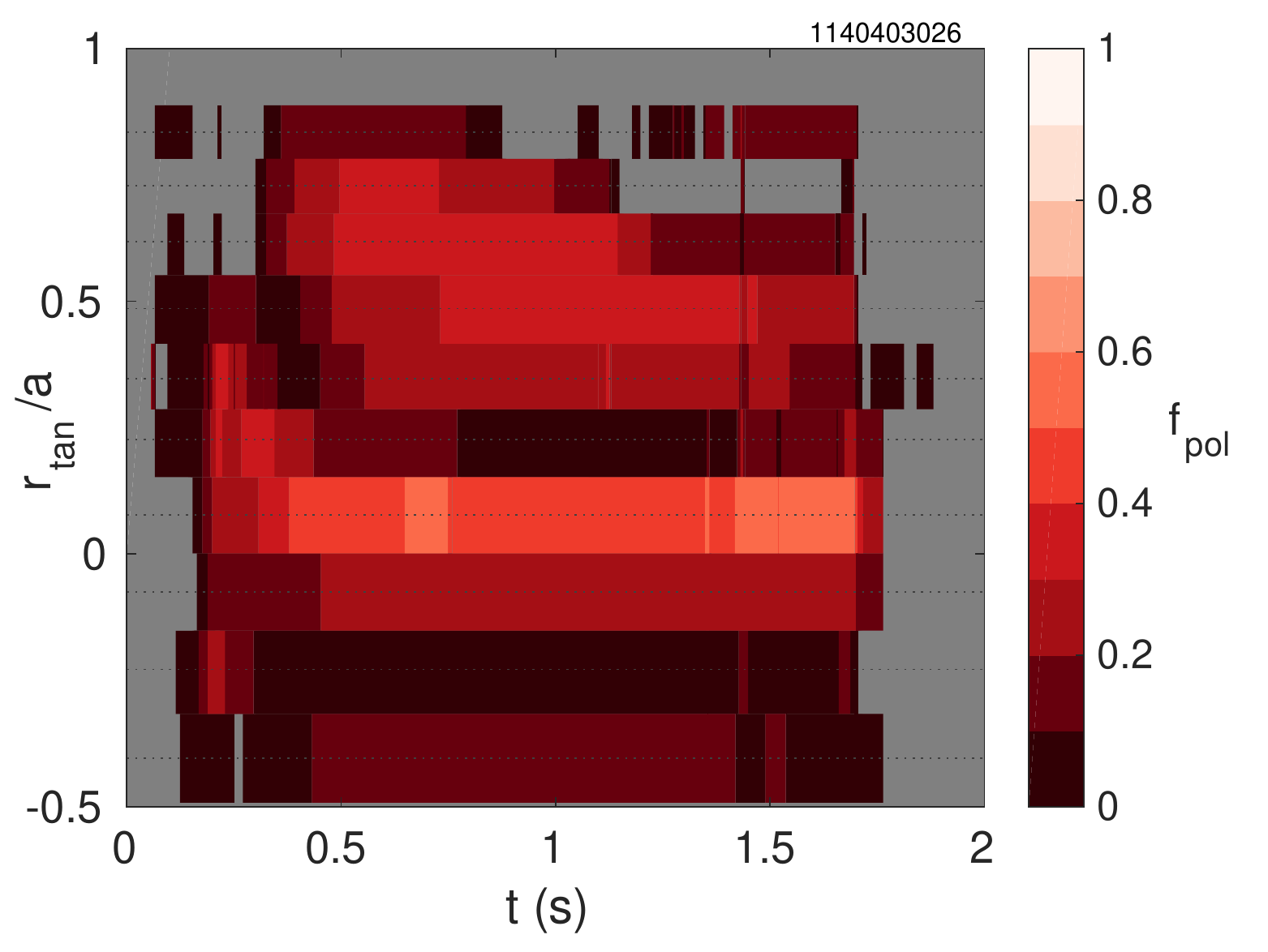}
        \caption{}
        \label{fig:pfExp}
    \end{subfigure}
    \caption{Experimentally-measured polarization (a) angle $\tpol$ and (b) fraction $\fpol$ from one Alcator C-Mod discharge, versus time and normalized tangency radius of the MSE channels (horizontal dotted lines). Time and spatial resolutions are $\Delta t \sim$~1~ms and $\Delta \rtan/a \sim$~0.1-0.2. Grey regions indicate signal below the noise floor.}
\end{figure}

Experimental measurements of $\tpol$ and $\fpol$ are shown in figures~\ref{fig:paExp} and \ref{fig:pfExp}, respectively, for all channels and times. Here, the time and spatial resolutions are $\Delta t \sim$~1~ms and $\Delta \rtan/a \sim$~0.1-0.2. In this work, we focus on the flattop portion of the discharge ($t \approx$~0.5-1.6~s) when plasma parameters are relatively stable. First consider the $\tpol$ data: Spatially, the $\pot$ transition occurs in the range $\rtan/a = -0.4$ to $-0.08$ (channel~1 to 3) and \redd{$\rtan/a = 0.23$ to 0.35 (channel~5 to 6).} Temporally, the most interesting $\tpol$ evolution is at $\rtan/a = -0.23$ (channel~2) which experiences $\pot$ transitions from $\tpol = 90\deg$ to $0\deg$ at $t \approx 0.7$~s and then back from $\tpol = 0\deg$ to $90\deg$ around $t \approx$~1.2-1.4~s. From our reference plot, figure~\ref{fig:paRef}, this implies that the dominant pitch angle of REs located within the channel~2 FOV decreases below $\tpc \approx 0.185$~rad for $t \approx$~0.7-1.4~s. Regarding experimental $\fpol$ measurements, a maximum value of $\fpol \approx 0.6$ is observed near the magnetic axis (channel~4), as expected. \redd{Note how the non-monotonic feature (i.e. the ``bump'') in $\fpol$ values at outer radii decreases in radial extent as the RE beam contracts in size.} This shrinking is confirmed \sout{by} \red{from} RE synchrotron images analyzed in~\cite{tinguely2018ppcf}.


\begin{figure}[h!]
    \centering
    \begin{subfigure}{\halfWidth}
        \centering
        \includegraphics[width=\textwidth]{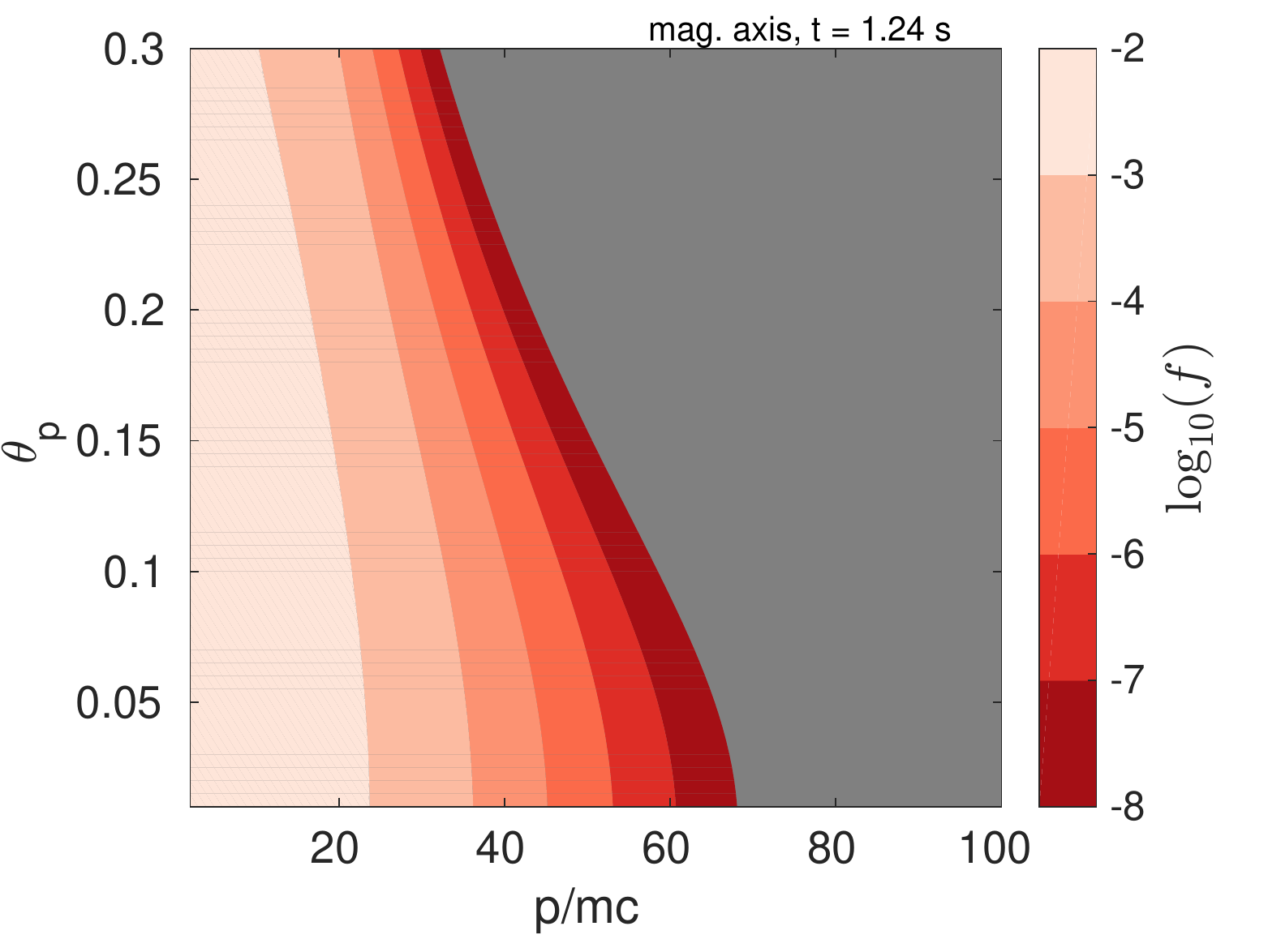}
        \caption{}
        \label{fig:CODE}
    \end{subfigure}
    \begin{subfigure}{\halfWidth}
        \centering
        \includegraphics[width=\textwidth]{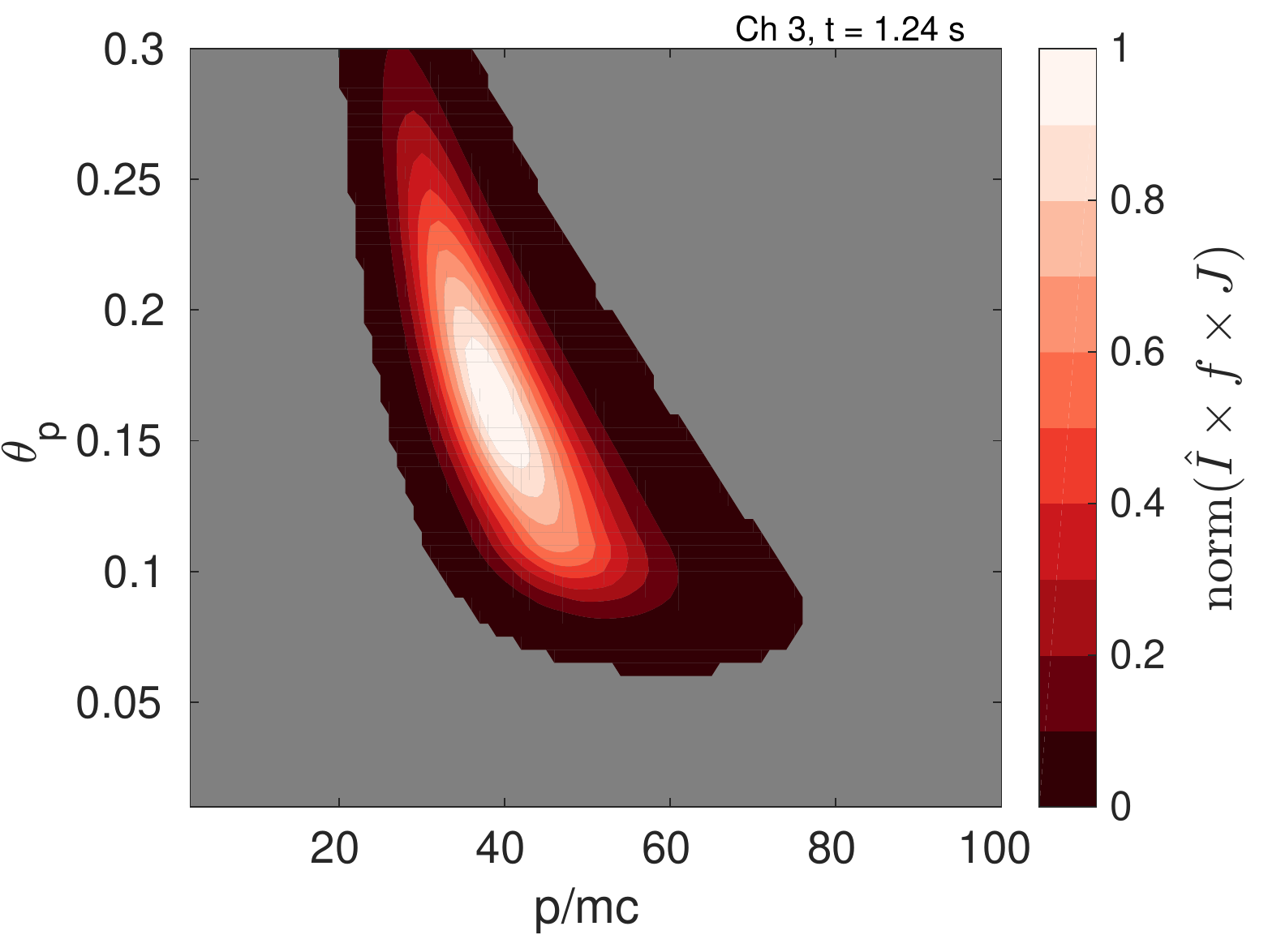}
        \caption{}
        \label{fig:IxfxJ}
    \end{subfigure}
    \caption{(a) A momentum space distribution function $f(p,\tp)$ (log scale) calculated by \CODE for plasma parameters at the magnetic axis. (b) The normalized convolution of $f$, the detector response function $\hat{I}(p,\tp)$ for channel~3, and Jacobian $J = p^2\sin\tp$. The location of peak detected emission is $p/mc \approx 40$ and $\tp \approx 0.16$~rad. \red{Grey regions indicate practically-undetectable regions of momentum space. ($t$~=~1.24~s)}}
    \label{fig:CODEAndIxfxJ}
\end{figure}

\begin{figure}[h!]
    \centering
    \begin{subfigure}{\halfWidth}
        \centering
        \includegraphics[width=\textwidth]{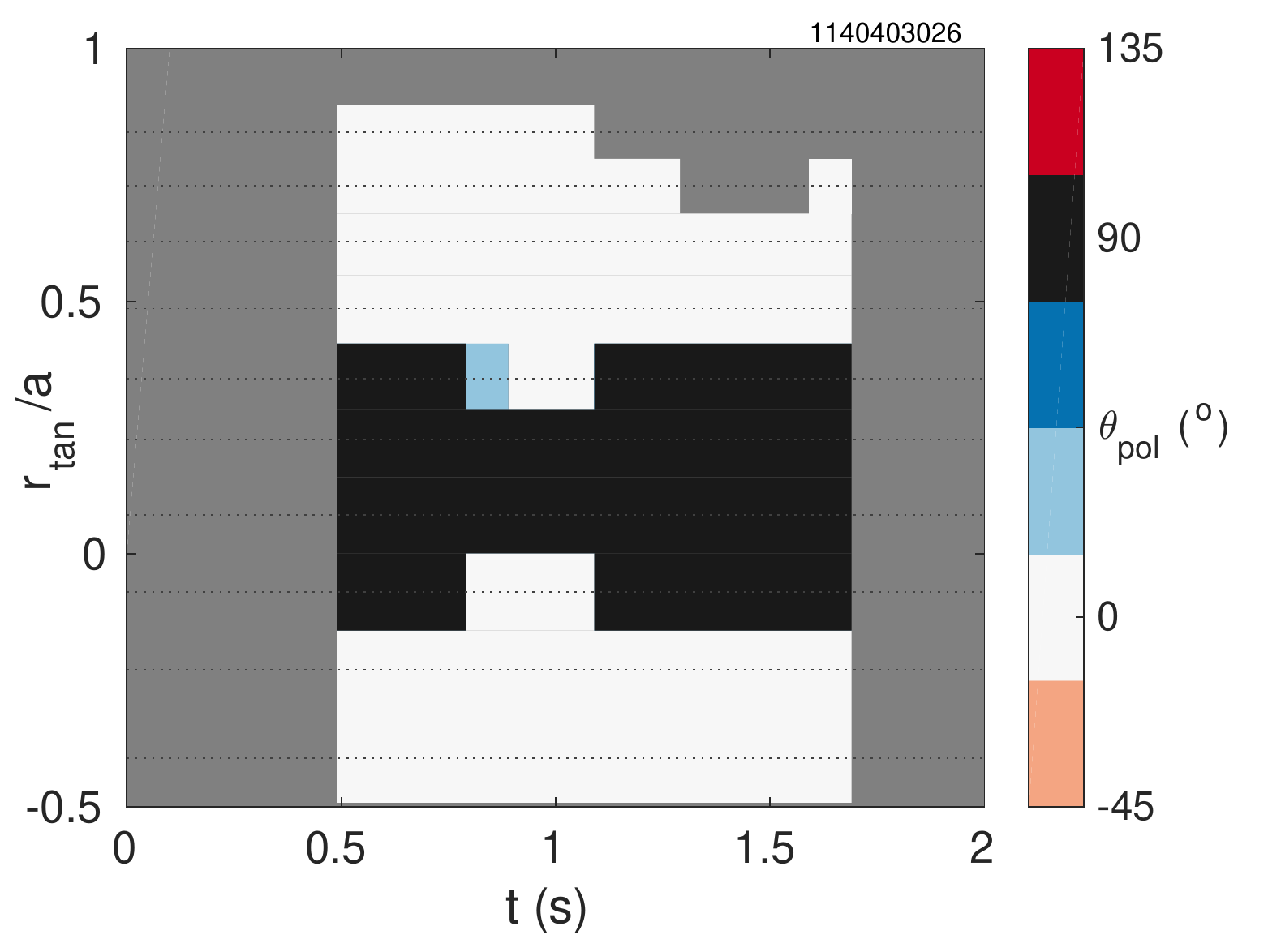}
        \caption{}
        \label{fig:paSOFT}
    \end{subfigure}
    \begin{subfigure}{\halfWidth}
        \centering
        \includegraphics[width=\textwidth]{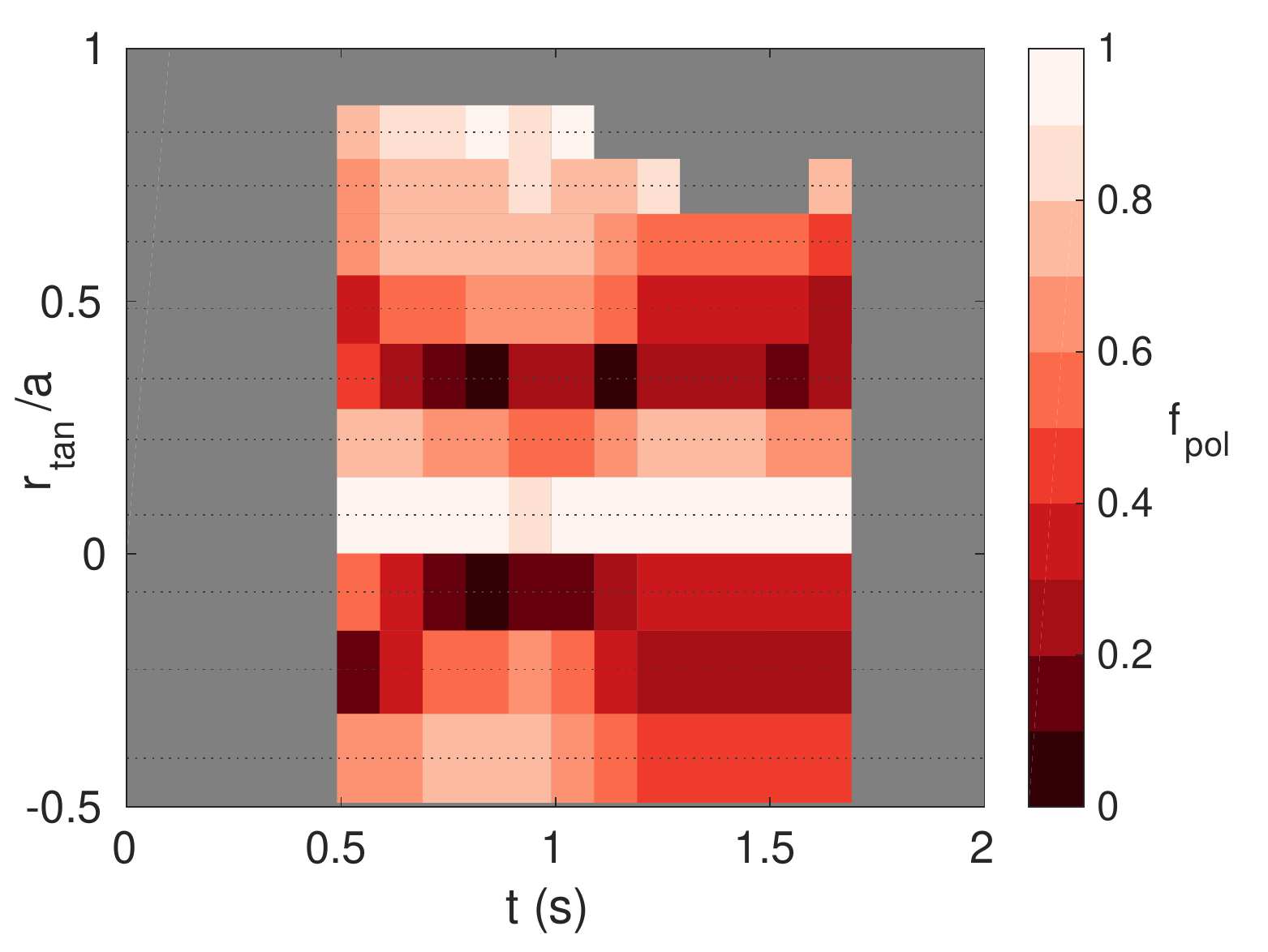}
        \caption{}
        \label{fig:pfSOFT}
    \end{subfigure}
    \caption{\SOFTCODE-predicted polarization (a) angle $\tpol$ and (b) fraction $\fpol$ for the Alcator C-Mod discharge of interest, versus time and normalized tangency radius of the MSE channels (horizontal dotted lines). Time and spatial resolutions are $\Delta t$~=~100~ms and $\Delta \rtan/a \sim$~0.1-0.2. Compare to figure~\ref{fig:paExp} and \ref{fig:pfExp}. \red{Grey regions indicate undetectable synthetic data ($\rtan/a \gtrsim 0.9$) and unexplored times ($t \lesssim 0.5$~s and $t \gtrsim 1.6$~s).}}
\end{figure}

The predicted \emph{synthetic} measurements shown in figures~\ref{fig:paSOFT} and \ref{fig:pfSOFT} were produced using the following methodology\footnote{This is the same approach as that followed in~\cite{tinguely2018ppcf}.}:

\begin{enumerate}
    \item Spatial profiles of experimental plasma parameters, like electron density and temperature from Thomson scattering and electric and magnetic fields from EFIT, were calculated for six locations throughout the plasma: at the magnetic axis \red{($q_{\rm axis} \approx 0.9$)} and on rational flux surfaces $q$~=~1, 4/3, 3/2, 2, and 3. A constant $\Zeff = 4$ was assumed.
    \item For each location, these parameters were input into the kinetic solver \emph{COllisional Distribution of Electrons} (\CODE)~\cite{landreman2014,stahl2016} to solve for the time-evolving electron momentum space distribution function. See figure~\ref{fig:CODE} for a sample time-slice. 
    \item \label{step:stitch} \redd{The RE \emph{phase} space distribution (i.e. in momentum and position space) was ``stitched'' together via a piecewise interpolation of the six momentum space distribution functions, with steps halfway between each flux surface as illustrated by the shaded regions in figure~\ref{fig:IvR}.} 
    \item The density profile $\nRE$ was inferred for the range $r/a \approx$~0.2-1 from experimental images of the synchrotron emission, as described in~\cite{tinguely2018ppcf}, and a Gaussian fit was used to extrapolate to $r/a = 0$. In general, it is observed that the \SOFT synthetic modeling of measurements is \emph{not} sensitive to the precise shape of \sout{the} $\nRE(R)$. 
    \item The entire phase space distribution was convolved with the \SOFT response functions (and Jacobian) for times $t$~=~0.54-1.64~s, with time step $\Delta t = 100$~ms limited by computation time. Integration over phase space then gives the synthetic \SOFTCODE results for $\tpol$ and $\fpol$, shown in figures~\ref{fig:paSOFT} and \ref{fig:pfSOFT}, respectively. 
\end{enumerate}

Figure~\ref{fig:IxfxJ} shows the convolution of the \CODE distribution function in figure~\ref{fig:CODE} with the detected intensity response function $\hat{I}$ of channel~3 ($\rtan/a = -0.08$). This highlights the region of momentum space ($p/mc \approx 40$ and $\tp \approx 0.16$~rad, in this case) which dominates the detected synchrotron measurement \red{of channel~3} and determines the synthetic measurements of $\tpol$ and $\fpol$ in figures~\ref{fig:paSOFT} and \ref{fig:pfSOFT}.

There are a few major takeaways when comparing experimental data with \SOFTCODE predictions. First, it is somewhat difficult to compare $\fpol$ data quantitatively. Although the experimental $\fpol$ measurements were calibrated, \SOFT does not account for the effects of background plasma light and reflections, which would decrease $\fpol$ and would be expected to have both spatial and temporal dependencies. However, there is relatively good \emph{qualitative} agreement between experimental and synthetic spatial profiles. Both have maximum $\fpol$ values near the magnetic axis ($\rtan/a = 0.08$ for channel~4), with minima on either side of this peak. In general, it is observed that \emph{increasing} RE pitch angles, \sout{e.g.} \red{for instance} through increased pitch angle scattering, would better match synthetic signals with the experimental measurements.


Second, consider channel~3 ($\rtan/a = -0.08$). In experiment, $\tpol \approx 90\deg$ is observed for all times (figure~\ref{fig:paExp}); however, from \SOFTCODE, $\tpol \approx 0\deg$ is predicted for $t$~=~0.64-0.84~s (figure~\ref{fig:paSOFT}). As seen in figure~\ref{fig:IvR}, channel~3 only ``sees'' REs near the magnetic axis, with a momentum space distribution modeled in \CODE using plasma parameters from the magnetic axis. Consulting our look-up plot in figure~\ref{fig:paRef}, experimental $\tpol$ data indicate that a significant fraction of REs must maintain $\tp \gtrsim 0.12$~rad for all times. This is achieved by the distribution function, from \CODE, at $t$~=~1.24~s shown in figure~\ref{fig:CODE}, which has emission dominated by REs with $\tp \approx$~0.16~rad, as seen in figure~\ref{fig:IxfxJ}. However, the \CODE-calculated pitch angle distribution falls below the threshold $\tpc \approx 0.12$~rad for $t$~=~0.64-0.84~s, which does not match experiment. 

Conversely, for channel~6 ($\rtan/a = 0.35$), experimental values of $\tpol = 0\deg$ are observed during the flattop current ($t > 0.5$~s), but \SOFTCODE predicts $\pot$ transitions at $t\approx$~0.64 and 0.84~s, similar to those predicted for channel~3 ($\rtan/a = - 0.08$). As seen in figure~\ref{fig:IvR}, channel~6 views REs in a radial range overlapping the momentum space distributions of REs on flux surfaces $q = 1$ and 4/3. From the look-up plot, figure~\ref{fig:paRef}, the bulk of the \emph{observed} RE population must have pitch angles $\tp \lesssim 0.12$~rad for all flattop times in order for measurements of $\tpol = 0\deg$ to be made. Therefore, the actual RE pitch angle distribution is inferred to be \emph{narrower} than that predicted by \CODE for times $t \approx$~0.5-0.7~s and 1.1-1.7~s.

Finally, consider once again channel~2 ($\rtan/a = -0.23$), for which an interesting time-evolution in experimental $\tpol$ measurements is observed in figure~\ref{fig:paExp}. \SOFTCODE, however, predicts that $\tpol = 0\deg$ for all times (see figure~\ref{fig:paSOFT}). As seen in figure~\ref{fig:IvR}, channel~2 views the radial range including the momentum space distributions of REs at the magnetic axis and on the flux surface $q = 1$. Referencing figure~\ref{fig:paRef}, it is seen that the RE pitch angle distribution from \CODE should \emph{broaden} (past the threshold $\tpc \gtrsim 0.185$~rad) for times $t \approx$~0.5-0.7~s and $\sim$1.3-1.7~s in order to improve agreement between synthetic and experimental data. In other words, the actual RE pitch angle distribution is inferred to be broader than that predicted by \CODE for those times. Note also from figure~\ref{fig:IvR} that channel~4 ($\rtan/a = -0.08$) views the same RE phase space distribution as channel~2. However, as previously mentioned, channel~4 provides little additional information, and \SOFTCODE synthetic data for channel~4 are consistent with experiment for all times.

To reiterate, the RE pitch angle distribution is inferred from experiment to be \emph{broader} than that predicted by \CODE in the radial region between the magnetic axis and surface $q \approx 1$, but \emph{narrower} between surfaces $q\approx 1$ and 4/3. It is not clear which physical mechanisms would cause an \emph{increase} of $\tp$ between the magnetic axis and surface $q \approx 1$ but a \emph{decrease} of $\tp$ in the region $q \approx$~1--4/3. One possible explanation is the interaction of the sawtooth instability with REs within and near the inversion radius $q = 1$. The partial reduction in $T_e$ sawteeth at the onset of the locked mode ($t \approx 0.7$~s) and complete suppression of sawteeth at $t \approx 1.1$~s correlate with the above times. This suggests that some of the differences between experimental and synthetic measurements are due to \CODE not accounting for spatial dynamics, such as radial transport. Note that while radial transport of REs near the surface $q = 2$ was inferred from synchrotron images in~\cite{tinguely2018ppcf}, RE dynamics within $q < 1$ were not observable; they are captured in the present work.

In general, it is seen that the synthetic and experimental $\tpol$ data match well for most channels and times. The observation that broadening the \CODE-predicted RE pitch angle distribution to improve agreement between synthetic and experimental $\fpol$ data indicates that additional pitch angle scattering mechanisms may not be captured in this model. The uniform $\Zeff = 4$ profile is assumed from previous experience with low density C-Mod discharges, but was not directly measured for this discharge. An increase in $\Zeff$ could cause increased pitch angle scattering. So, too, could the magnetic fluctuations related to the locked mode or perhaps another RE-induced kinetic instability. Because pitch angle scattering \emph{increases} as particle energies \emph{decrease}, an unaccounted power loss mechanism might also explain the results. These mechanisms could be isolated and better diagnosed in future experiments.


\section{Summary}\label{sec:summary}

This paper presented the first experimental analysis of polarized synchrotron emission from relativistic runaway electrons (REs) in a tokamak plasma. These RE experiments were performed during low density, Ohmic, elongated and diverted plasma discharges in the Alcator C-Mod tokamak. Due to the high magnetic field on-axis $\Bo$~=~5.4~T, significant levels of visible synchrotron radiation were detected by the ten-channel Motional Stark Effect (MSE) diagnostic, which measured spatial profiles of the intensity $L$ and fraction $\fpol$ of detected light which was linearly-polarized, as well as the polarization angle $\tpol$. Data from 28 plasma discharges (corresponding to over one thousand time-points), during which synchrotron-producing REs were generated, indicated that measurements of $\tpol$ and $\fpol$ are strongly dependent on the detector and magnetic geometries. An interesting spatial feature, a $\pot$ transition in $\tpol$ first predicted by Sobolev \cite{sobolev2013}, was also observed in the experimental data.


The modeling of polarized synchrotron emission via Stokes parameters and its implementation in the versatile synthetic diagnostic \SOFT \cite{hoppe2018} were discussed. The $\pot$ transition in $\tpol$ was intuitively explained by the ``cone'' approximation of RE synchrotron emission: Because the electric field $\vE$ is approximately proportional to the Lorentz acceleration $\vv \times \vB$, horizontal or vertical polarization of synchrotron radiation is observed when the detector ``sees'' the top/bottom or sides, respectively, of the emission cone. Thus, a critical RE pitch angle $\tpc$, at which the $\pot$ transition occurs, was calculated from the detector inclination and field-of-view opening angle, as well as local magnetic field pitch. Green's functions from \SOFT---which give a synthetic $\tpol$ or $\fpol$ value when integrated with a RE phase space distribution---confirmed the existence of this $\tpc$, at which the polarization fraction $\fpol$ was also expected and confirmed to be minimal. Furthermore, it was found that $\tpol$ and $\fpol$ measurements depend little on the RE energy or density profile. From the Green's functions, ``look-up'' plots (see figure~\ref{fig:Ref}) were created from which experimental values of $\tpol$ or $\fpol$ could be used to constrain the pitch angle $\tp$ of the REs which dominate the synchrotron radiation measurement.

The spatiotemporal evolutions of $L$, $\tpol$, and $\fpol$ signals were explored in detail for one C-Mod discharge. Experimentally-measured plasma parameters were input into the kinetic solver \CODE \cite{landreman2014,stahl2016} to calculate the momentum space distribution of REs located at the magnetic axis and on flux surfaces $q$~=~1, 4/3, 3/2, 2, and 3. These were input into \SOFT to compute synthetic signals which could then be compared to experiment. In general, synthetic $\tpol$ measurements from \SOFT were found to match experimental values for most channels and times. Disagreements were found in channels viewing REs near the magnetic axis and surfaces $q$~=~1 and 4/3. When compared to the predicted RE pitch angle distribution from \CODE, it was inferred that the actual RE pitch angle distribution was (i) dominated by \emph{larger} pitch angles for REs located within $q \lesssim$~1, but (ii) dominated by \emph{smaller} pitch angles for REs approximately \sout{between surfaces} \red{within} $q \in$~[1,~4/3]\red{; this could indicate} \sout{possibly indicating} an interaction of the sawtooth instability and/or locked mode with REs that was not captured by \CODE. Moreover, it was seen that \emph{increasing} RE pitch angles could give better agreement between synthetic and experimental $\fpol$ measurements; therefore, additional pitch angle scattering mechanisms, e.g. from kinetic instabilities, may need to be incorporated in future analyses.

In summary, this paper has shown that the polarization information of synchrotron emission can be used as a novel diagnostic of the RE pitch angle distribution. It is conceivable that these measurements could be made by the MSE systems of other tokamaks if significant synchrotron light is produced in the MSE wavelength range. Moreover, the interplay of the RE pitch angle, magnetic geometry, and polarization of synchrotron light opens an opportunity to probe the RE current density profile through its contribution to the poloidal magnetic field. This could be especially important for post-disruption RE beams, which were not achievable in C-Mod. Finally, this work motivates an exploration of better RE diagnostics; for instance, one could combine the benefits of a camera, spectrometer, and polarimeter to gather spatial, energy, and pitch angle information, respectively, in the pursuit of constraining and inverting the entire RE phase space distribution.


\section*{Acknowledgements}

The authors thank O. Embr\'{e}us, J. Decker, G. Papp, and Y. Peysson for fruitful discussions, as well as the entire Alcator C-Mod team. This work was supported by US DOE Grant DE-FC02-99ER54512, using Alcator C-Mod, a DOE Office of Science User Facility; Vetenskapsr\aa det (Dnr 2014-5510); and the European Research Council (ERC-2014-CoG grant 647121).  





\section*{References}

\end{document}